\documentclass[tradiabstract]{aa}  

\usepackage{graphicx}
\usepackage{lscape}
\usepackage{amsmath}
\usepackage{txfonts}
\usepackage{url}
\usepackage{natbib}
\usepackage[usenames,dvipsnames,svgnames,table]{xcolor}
\usepackage{ulem}
\usepackage[utf8]{inputenc}
\usepackage{siunitx}

\newcommand{\kms}{km~s$^{-1}$}
\newcommand{\Msun}{M$_\odot$}
\newcommand{\Rsun}{R$_\odot$}

\newcommand{\Mjup}{M$_J$}
\newcommand{\Rjup}{R$_J$}
\newcommand{\Ha}{H${\alpha}$}
\newcommand{\Hb}{H${\beta}$}
\newcommand{\Hg}{H${\gamma}$}
\newcommand{\Hd}{H${\delta}$}
\newcommand{\He}{H${\epsilon}$}
\newcommand{\Hz}{H${\zeta}$}

\begin{document} 

% Title Page
\title{Mass-loss rate and local thermodynamic state of the KELT-9~b thermosphere from the hydrogen Balmer series\thanks{Based on observations made at TNG (Telescopio Nazionale Galileo) telescope with the HARPS-N spectrograph under program A35DDT4 and OPTICON 2018A/038.}}

\author{A.~Wyttenbach\inst{1,2}\thanks{Fellow of the Swiss National Science Foundation (SNSF)}, P.~Molli\`ere\inst{1,3}, D.~Ehrenreich\inst{4}, H.~M.~Cegla\inst{4}\thanks{CHEOPS Fellow, SNSF NCCR-PlanetS}, V.~Bourrier\inst{4}, C.~Lovis\inst{4}, L.~Pino\inst{5}, R.~Allart\inst{4}, J.~V.~Seidel\inst{4}, H.~J.~Hoeijmakers\inst{4,6}, L.~D.~Nielsen\inst{4}, B.~Lavie\inst{4}, F.~Pepe\inst{4}, and X.~Bonfils\inst{2}, and I.~A.~G.~Snellen\inst{1}}
\authorrunning{Wyttenbach et al.}
\titlerunning{Hydrogen Balmer series in the KELT-9~b thermosphere}
\institute{Leiden Observatory, Leiden University, Postbus 9513, 2300 RA Leiden, The Netherlands\label{inst1}
\and Universit\'e Grenoble Alpes, CNRS, IPAG, 38000 Grenoble, France
\and Max-Planck-Institut f\"ur Astronomie, K\"onigstuhl 17, D-69117 Heidelberg, Germany
\and Geneva Observatory, University of Geneva, ch. des Maillettes 51, CH-1290 Versoix, Switzerland
\and Anton Pannekoek Institute for Astronomy, University of Amsterdam, Science Park 904, 1098 XH Amsterdam, The Netherlands
\and University of Bern, Center for Space and Habitability, Sidlerstrasse 5, CH-3012, Bern, Switzerland
\\
\email{aurelien.wyttenbach@univ-grenoble-alpes.fr}}

\date{Received 16 December 2019; accepted 25 April 2020} 
\abstract
  % aims heading (mandatory)
{KELT-9~b, the hottest known exoplanet with $\mathrm{T_{eq}\sim4400}$~K, is the archetype of a new planet class known as ultra-hot Jupiters. These exoplanets are presumed to have an atmosphere dominated by neutral and ionized atomic species. In particular, \Ha\ and \Hb\ Balmer lines have been detected in the KELT-9~b upper atmosphere, suggesting that hydrogen is filling the planetary Roche lobe and escaping from the planet. In this work, we detected $\delta$ Scuti-type stellar pulsation (with a period $P_{\rm puls}=7.54\pm0.12$~h) and studied the Rossiter-McLaughlin effect (finding a spin-orbit angle $\lambda=\ang{-85.01}\pm\ang{0.23}$) prior to focussing on the Balmer lines (\Ha\ to \Hz) in the optical transmission spectrum of KELT-9~b. Our HARPS-N data show significant absorption for \Ha\ to \Hd. The precise line shapes of the \Ha, \Hb, and \Hg\ absorptions allow us to put constraints on the thermospheric temperature. Moreover, the mass loss rate, and the excited hydrogen population of KELT-9~b are also constrained, thanks to a retrieval analysis performed with a new atmospheric model. We retrieved a thermospheric temperature of $T=13\,200^{+800}_{-720}$~K and a mass loss rate of $\dot{\rm M}=10^{12.8\pm0.3}$~g\,s$^{-1}$ when the atmosphere was assumed to be in hydrodynamical expansion and in local thermodynamic equilibrium (LTE). Since the thermospheres of hot Jupiters are not expected to be in LTE, we explored atmospheric structures with non-Boltzmann equilibrium for the population of the excited hydrogen. We do not find strong statistical evidence in favor of a departure from LTE. However, our non-LTE scenario suggests that a departure from the Boltzmann equilibrium may not be sufficient to explain the retrieved low number densities of the excited hydrogen. In non-LTE, Saha equilibrium departure via photo-ionization, is also likely to be necessary to explain the data.}
%    {}
%    {}
% methods heading (mandatory)
%    {}
%results heading (mandatory)
%    {}
%    {}


\maketitle

%%%%%%%%%%%%%%%%%%%%%%%%%%%%%%%%%%%%%%%%%%%%%%%%%%%%%%%%%%%%%%%
\section{Introduction}
%%%%%%%%%%%%%%%%%%%%%%%%%%%%%%%%%%%%%%%%%%%%%%%%%%%%%%%%%%%%%%%

Observations of exoplanet atmospheres allow us to characterize several of these remote worlds. Indeed, the measurement of planetary spectra are rich in information about these planets' physical and chemical conditions. Thus, the study of exoplanetary atmospheres are motivated by the possibility that they can provide crucial hints about their origin and evolution.

When observing the primary transit of an exoplanet as a function of the wavelength, we measure a transmission spectrum. The transmission spectrum of an exoplanet bears the signature of its atmospheric constituents. The possibility of detecting atomic or molecular species through transmission spectroscopy was theorized early on \citep{Seager2000,Brown2001}. Since giant close-in planets \citep{Mayor1995}, with hot and extended atmospheres, are easy targets to probe in transmission, early observations of hot Jupiter atmospheres were achieved with sodium detections \citep{Charbonneau2002,Redfield2008,Snellen2008}. Recent results suggest that high-resolution (typically $\lambda/\Delta\lambda\sim10^5$) transmission spectra of hot Jupiters probe their thermospheres, a key region of the upper atmosphere, in order to understand irradiated planets and allow us to achieve precise temperature and wind measurements because the lines (e.g., the \ion{Na}{i} D doublet, the \ion{K}{i} D doublet, the \ion{H}{i} Balmer lines H${\alpha,\beta,\gamma}$, the \ion{He}{i} ($2^3$S) triplet) are resolved and studies benefit from large signals \citep[e.g.,][]{Wyttenbach2015,Wyttenbach2017,Casasayas-Barris2017,Casasayas-Barris2018,Casasayas-Barris2019,Khalafinejad2017,Khalafinejad2018,Seidel2019,Yan2018,Cauley2018,Allart2018,Allart2019,Nortmann2018,Salz2018,Alonso-Floriano2019b,Keles2019,Chen2020}. Another landmark achievement obtained in the optical domain is the detection, with cross-correlation, of iron and other metals in the atmosphere of several ultra-hot Jupiters \citep{Hoeijmakers2018,Hoeijmakers2019,Borsa2019,Bourrier2020a,Cabot2020,Gibson2020,Nugroho2020,Stangret2020,Ehrenreich2020}.

\begin{figure}[t!]
\centering
\includegraphics[width=0.47\textwidth]{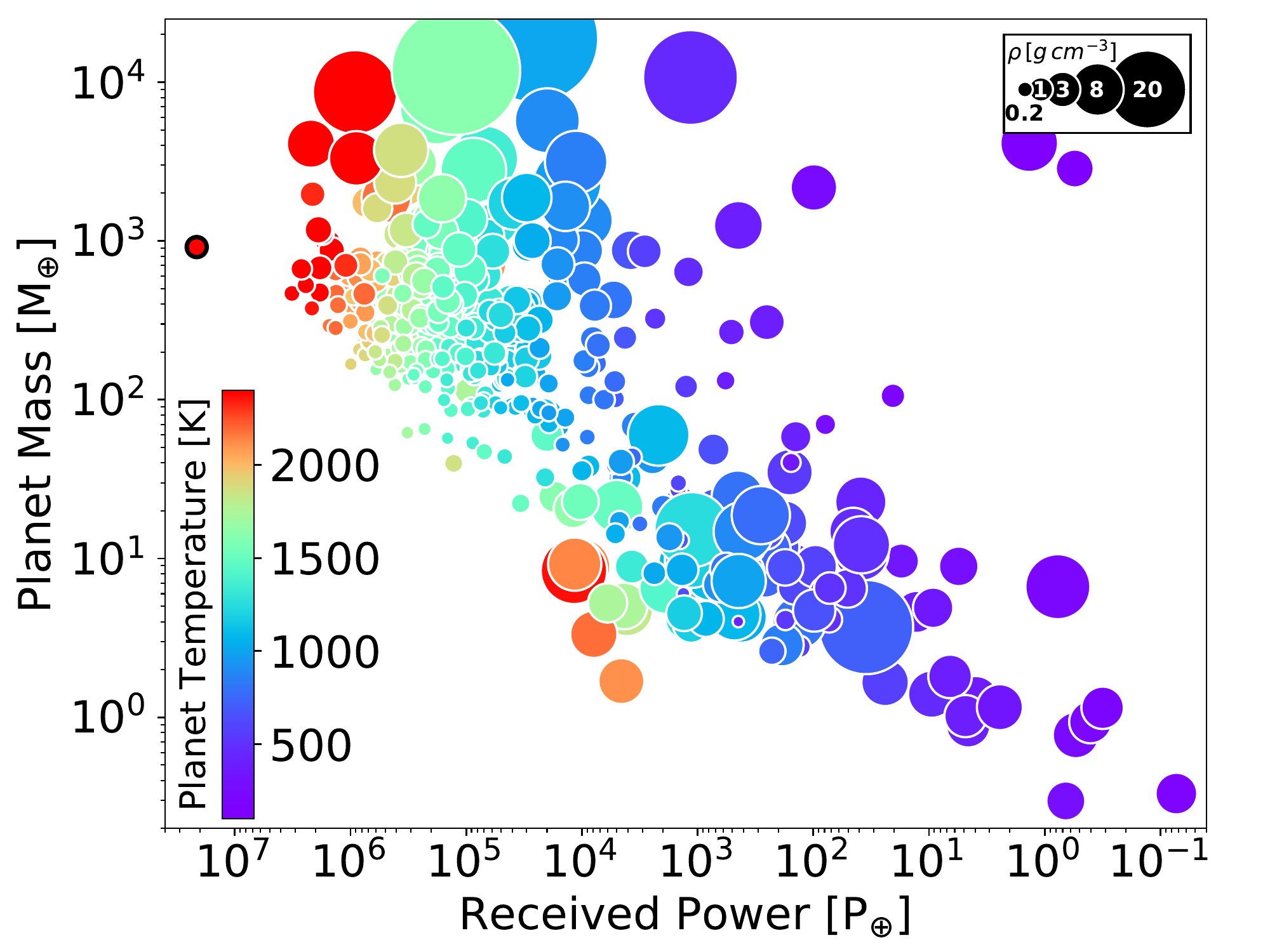}
\caption{Distribution of planetary masses as a function of the total power received by the planet from the incident bolometric stellar radiation, expressed in units of power received on Earth at the top of the atmosphere (P$_{\oplus}=174~\mathrm{PW}=1.74\times10^{24}~\mathrm{erg~s^{-1}}$). The marker size scales with the planet density, and the color with the planetary equilibrium temperature. Physical properties of exoplanets were extracted from the Extrasolar Planets Encyclopedia (exoplanet.eu) in December 2019. The lower left corner represents the evaporation desert. KELT-9 (HD~195689), encircled in black, stand out of all other planetary population in term of its received power and equilibrium temperature.}
\label{planet_diagram}
\end{figure}

Ultra-hot Jupiters are the hottest and the most irradiated hot Jupiters (with a planetary equilibrium temperature above 2,000-2,500~K, Fig.~\ref{planet_diagram}). Well-known members of that category include KELT-9~b \citep[][the hottest exoplanet discovered so far around a main-sequence star, with $\mathrm{T_{eq}\gtrsim4,000}$~K, see Table~\ref{tab:system}]{Gaudi2017}, the newly discovered MASCARA-4~b/bRing-1~b \citep[][]{Dorval2019}, as well as targets often observed due to their bright host stars, namely, WASP-189~b, WASP-33~b, MASCARA-1~b, WASP-12~b,  WASP-103~b, WASP-18~b, WASP-121~b, KELT-20~b/MASCARA-2~b, HAT-P-7~b, WASP-76~b, etc. Hot exoplanet atmospheres are expected to be near chemical equilibrium from $T\gtrsim2,500$~K \citep{Miguel2014,Lothringer2018,Kitzmann2018}. Under such assumptions, planet atmospheres are expected to be cloud-free on the day-side, to have strong inverted temperature-pressure profiles, and to be nearly barren of molecules, apart from H$_2$, CO and possible traces of H$_2$O, TiO and OH. Because of thermal dissociation and ionization, mostly neutral and ionized atomic hydrogen and helium, along with such metals as iron, sodium, magnesium, titanium, etc., are expected to be found \citep{Kitzmann2018,Lothringer2018,Lothringer2019,Mansfield2019}. In this type of atmosphere, the important source of H$^-$ bound-free and free-free opacities acts as a dominant continuum opacity \citep{Arcangeli2018,Parmentier2018,Bourrier2019a}. These predictions can change due to heat transport and circulation, depending on whether we observe the day or the night side, or the terminator \citep{Parmentier2018,Bell2018}. For example, in emission spectroscopy, these features of ultra-hot Jupiters are mixed up with the H$_2$O and TiO emission features (that may be absent due to dissociation), which make emission spectra appear as a continuum \citep[see][]{Parmentier2018,Arcangeli2018,Mansfield2018,Kreidberg2018}. In transmission spectroscopy, the probed pressures are lower, thus increasing the presence of atomic species because there is a higher rate of thermal dissociation. These higher rates are produced even without taking disequilibrium processes into account. In particular, the effect of the H$^-$ opacity increases, playing a more crucial role in the modeled transmission spectra \citep{Kitzmann2018}.

\begin{table}
\caption{Adopted values for the orbital and physical parameters of the KELT-9 system.}
\begin{center}
% \tablefootmark{\textasteriskcentered}
% \tablefootmark{\dag}
% \tablefootmark{\ddag}
% \tablefootmark{\S}
% \tablefootmark{\P}
\begin{tabular}{lc}
\hline
%\rule[0mm]{0mm}{5mm}Parameters & Value\rule[0mm]{0mm}{3mm}\\
Parameter & Value\tablefootmark{\textasteriskcentered} \\
\hline
\\
\multicolumn{2}{l}{Star: KELT-9 (HD~195689)}\\
$V$ mag                 & 7.55\\
Spectral type           & B9.5-A0\,{\sc v}\\
$M_{\star}$             & $1.978\pm0.023$~\Msun\tablefootmark{\dag}\\
$R_{\star}$             & $2.178\pm0.011$~\Rsun\tablefootmark{\dag}\\
$\rho_{\star}$             & $0.2702\pm0.0029$~g~cm$^{-3}$\tablefootmark{\dag}\\
$T_\mathrm{eff}$        & $9\,600\pm400$~K\tablefootmark{\S}\\
$\log g_{\star}$                & $4.058\pm0.010$~cgs\tablefootmark{\dag}\\
$v \sin i_{*}$ Spectro          & $112.60\pm0.04$~\kms\tablefootmark{\ddag}\\
$v \sin i_{*}$ RM (SB)         & $122.5\pm0.6$~\kms\tablefootmark{\ddag}\\
$v \sin i_{*}$ RM (DR)         & $116.9\pm1.8$~\kms\tablefootmark{\ddag}\\
\\
\multicolumn{2}{l}{Transit}\\
$T_0$                   & $2457095.68572\pm0.00014~\mathrm{BJD_{tdb}}$\\
$P$                     & $1.4811235\pm0.0000011$~d\\
$(R_p/R_\star)^2$       & $0.00677\pm0.000072$\\
$\Delta T_\mathrm{1-4}$ & $3.91584\pm0.00384$~h\\
$b$                     & $0.177\pm0.014~R_{\star}$\\
$a$                     & $0.0346\pm0.001$~au\\
$i$                     & $86.79\degr\pm0.25\degr$\\
$\lambda$	 SB	& $-85.65\pm0.09\degr$\tablefootmark{\ddag}\\
$\lambda$ DR	& $-85.01\pm0.23\degr$\tablefootmark{\ddag}\\
linear LD $u_1$               & $0.35$ (adopted)\\
\\
\multicolumn{2}{l}{Radial velocity}\\
$K_1 = K_{\star}$                   & $276\pm79$~m~s$^{-1}$\\
$K_2 = K_p$                   & $234.24\pm0.90$~\kms\tablefootmark{\dag}\\
$e$                     & 0 (adopted)\\
$\omega$                & $90\degr$ (adopted)\\
$\gamma$                & $-17\,740\pm40$~m~s$^{-1}$\tablefootmark{\dag}\\
\\
\multicolumn{2}{l}{Planet: KELT-9~b (HD~195689b)}\\
$M_p$                   & $2.44\pm0.70$~\Mjup\tablefootmark{\dag}\\
$R_p$                   & $1.783\pm0.009$~\Rjup\tablefootmark{\dag}\\
$\rho_p$                  & $0.53\pm0.15$~g~cm$^{-3}$\tablefootmark{\ddag}\\
$g_\mathrm{surf}$       & $1990\pm600$~cm~s$^{-2}$\tablefootmark{\ddag}\\
$T_\mathrm{eq}$         & $4360\pm200$~K\tablefootmark{\P}\\ 
$H_0$                     & $1400\pm400$~km\tablefootmark{\P}\\
$2R_pH_0/R_\star^2$       & $150\pm50$~ppm\tablefootmark{\P}\\
\hline
\end{tabular}
\end{center}
\tablefoot{}
\tablefoottext{\textasteriskcentered}{The parameters and errors are initially taken or computed from the extended data table 3 in \citet{Gaudi2017}.}
\tablefoottext{\dag}{Updated values from \citet{Hoeijmakers2019}.}
\tablefoottext{\ddag}{Updated values from this work.}
\tablefoottext{\S}{Updated values from \citet{Borsa2019}.}
\tablefoottext{\P}{For the planetary equilibrium temperature $T_\mathrm{eq}$ and the lower atmosphere scale-height $H_0$, we assume an albedo $A=0$, a redistribution factor $f=1$ and a mean molecular weight $\mu_{\mathrm{atm}}=1.3$ (atomic H and He atmosphere). We also consider the planet distance to the stellar surface to be $o = a-R_{\star}$.}
\label{tab:system}
\end{table}

\begin{table*}
\caption{Log of HARPS-N observations.}
\begin{center}
\begin{tabular}{lcccccccccc}
\hline
\rule[0mm]{0mm}{5mm} & Date & $\#$ Spectra\tablefootmark{\textasteriskcentered} & Exp.& Airmass\tablefootmark{\dag} & Seeing &  $S/N$ & $S/N$ & $S/N$ & $S/N$ & $S/N$\\
& & & [s] & & & 400 nm\tablefootmark{\ddag} & 600 nm\tablefootmark{\ddag} & \Ha\ core\tablefootmark{\S} & \Hb\ core\tablefootmark{\S}	& \Hg\ core\tablefootmark{\S}\\
\hline
$\mathrm{Night\ 1}$ & 2017-07-31 & 49 (22/27) & 600 & 1.5--1.02--1.7 & -- & 60--131 & 98--189 & 60--120 & 64--128 & 51--105\\
$\mathrm{Night\ 2}$ & 2018-07-20 & 46 (22/24) & 600 & 1.7--1.02--1.4 & -- & 27--79 & 70--160 & 44--100 & 64--114 & 41--79\\
\hline
\end{tabular}
\end{center}
\tablefoot{}
\tablefoottext{\textasteriskcentered}{In parentheses: the number of spectra taken during the transit and outside the transit, respectively.}
\tablefoottext{\dag}{The three figures indicate the airmass at the start, middle and end of observations, respectively.}
\tablefoottext{\ddag}{The range of signal-to-noise ($S/N$) ratio per pixel extracted in the continuum near 400~nm and 600~nm.}
\tablefoottext{\S}{The range of $S/N$ ratio in the line cores of the \Ha, \Hb\ and \Hg\ lines.}
\label{tab:log}
\end{table*}

Recent observations of KELT-9~b in transmission spectroscopy bring in key aspects for our understanding of ultra-hot Jupiters. The detections of \ion{Fe}{i}, \ion{Fe}{ii}, \ion{Ti}{ii} \citep[but not of \ion{Ti}{i};][]{Hoeijmakers2018} are confirming the predicted presence of neutral and ionized species in ultra-hot Jupiters. In \citet{Cauley2018,Hoeijmakers2019,Yan2019,Turner2020}, more neutral and ionized atomic species are reported, such as \ion{Na}{i}, \ion{Ca}{i}, \ion{Ca}{ii}, \ion{Mg}{i}, \ion{Cr}{ii}, \ion{Sc}{ii}, and \ion{Y}{ii}. In KELT-9~b, ionized species are more abundant than neutral species, which is expected for such high temperatures. However, the absorptions measured for ionized species are not reproduced by the models \citep{Hoeijmakers2019,Yan2019}. Those measurements probe altitudes between 1.01 to 1.1 planetary radii (1--15 scale-height or $10^{-3}$--$10^{-6}$ bar), thus, in the lower thermosphere. A complete exploration of species absorbing at optical wavelengths in ultra-hot Jupiter is very constraining in terms of aeronomic conditions. Indeed, the additional detections of \Ha\ by \citet{Yan2018,Turner2020}, and of \Ha, \Hb\ by \citet{Cauley2018},  also bring in important clues about the KELT-9 system because measuring absorption from specific strong opacity lines, such as the sodium \ion{Na}{i} D doublet or the hydrogen Balmer lines, allow us to probe higher layers ($\gtrsim1.2$ planetary radii) in the thermosphere \citep{Christie2013,Heng2015,Huang2017}. So far \Ha\ has only been observed in six exoplanets, four of which are classed as ultra-hot Jupiters \citep{Jensen2012,Jensen2018,Cauley2015,Cauley2016,Cauley2017a,Cauley2017b,Cauley2017c,Cauley2018,Barnes2016,Casasayas-Barris2018,Casasayas-Barris2019,Cabot2020,Chen2020}. The detection of an optically thick \Ha\ absorption in the atmosphere of KELT-9~b, which takes place at up to $\sim$1.6 planetary radii (at $\sim$10,000~K in the thermosphere), provides hints that the hydrogen is filling up the planetary Roche lobe (1.95 R$_p$) and is escaping from the planet at a rate of $\sim$$10^{12}$~g\,s$^{-1}$. This confirms that such planets undergo a process of evaporation \citep{Sing2019}. However, it is important to get a more precise idea of the mass loss rate of such objects to understand the role of the stellar irradiation and its processes of interaction with the planet atmosphere \citep{Fossati2018,Allan2019,GarciaMunoz2019}.

This paper reports on the results of the Sensing Planetary Atmospheres with Differential Echelle Spectroscopy (SPADES) program \citep{Bourrier2017a,Bourrier2018a,Hoeijmakers2018,Hoeijmakers2019}. This program, performed with the HARPS-N spectrograph, is a twin of the Hot Exoplanetary Atmospheres Resolved with Transit Spectroscopy (HEARTS) program based on the use of the HARPS spectrograph \citep{Wyttenbach2017,Seidel2019,Bourrier2020a}. Throughout this work, we introduce the python \texttt{CHESS} code (the CHaracterization of Exoplanetary and Stellar Spectra code) and its sub-modules, which allow us to measure, separate, and model the stellar and exoplanet spectra.

The HARPS-N transit observations\footnote{This data set is the same as in \citet{Hoeijmakers2018,Hoeijmakers2019}. Those previous studies focused on metal detections (species between atomic numbers 3 and 78), while our present study focuses only on the hydrogen lines, making it a complementary analysis. We note that no Helium ($\lambda 10830\AA$) has been found in \citet{Nortmann2018}.} are described in Sect.~\ref{Sec_Obs}. In Sect.~\ref{Sec_newTS}, we propose a comprehensive analysis of spectroscopic transit data and present an improved method to extract transmission spectra. Furthermore, we show a stellar pulsation analysis and a Rossiter-McLaughlin analysis, and we present the Balmer lines signatures we found in the transmission spectrum of KELT-9~b. In Sect.~\ref{Sec_inter}, we present a new model and retrieve fundamental atmospheric properties from the high-resolution data observations. In Sec~\ref{Sec_Discuss}, we discuss the constraints we find on the KELT-9b atmospheric temperature, mass loss rate, and local thermodynamic properties. We present our conclusions in Sect. 6.

%%%%%%%%%%%%%%%%%%%%%%%%%%%%%%%%%%%%%%%%%%%%%%%%%%%%%%%%%%%%%%%
\section{The HARPS-N dataset}\label{Sec_Obs}
%%%%%%%%%%%%%%%%%%%%%%%%%%%%%%%%%%%%%%%%%%%%%%%%%%%%%%%%%%%%%%%

\begin{figure*}[t!]
\centering
\includegraphics[width=0.97\textwidth]{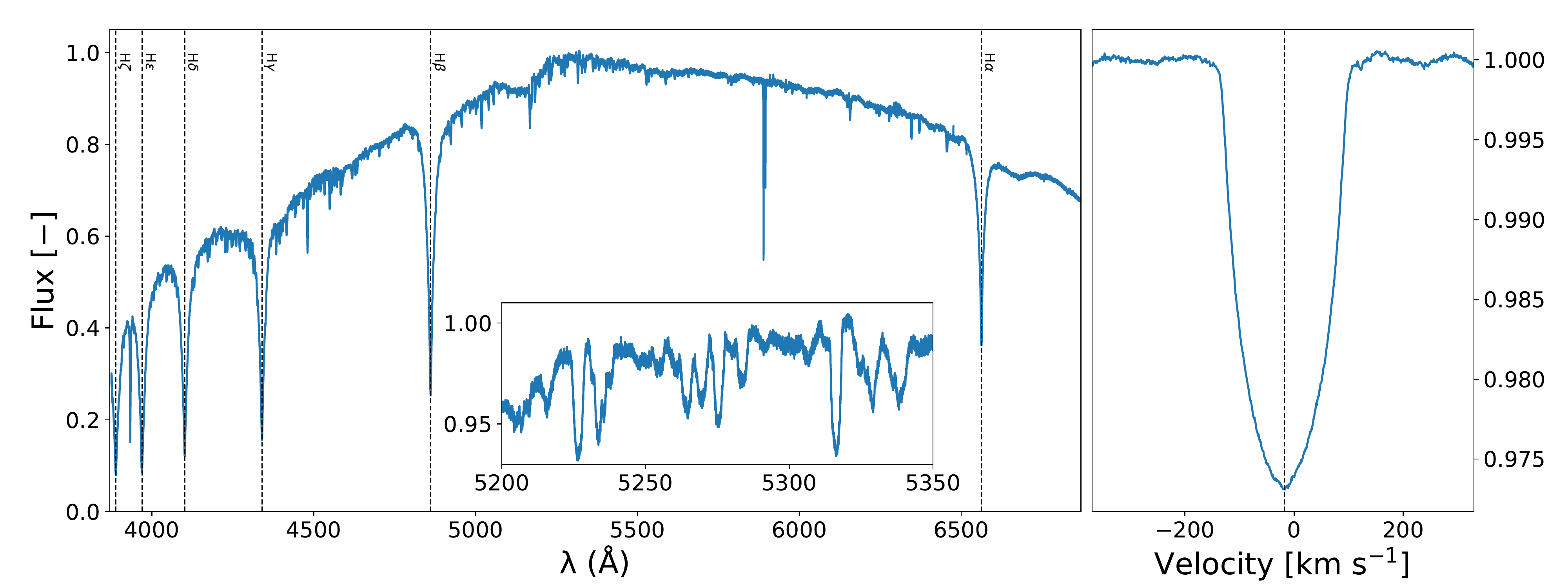}
\caption{Left panel: Normalized out-of-transit master spectrum of KELT-9 in the 387--687~nm optical region. The stellar A0 spectral type makes the spectrum dominated by strong Balmer lines, though we notice the presence of numerous shallow metal lines, as shown in the inset. The Balmer lines (H${\alpha,\beta,\gamma,\delta,\epsilon,\zeta}$), that are covered by the HARPS-N spectrograph wavelength range, are identified by the vertical dashed lines. Right Panel: Normalized out-of-transit master cross-correlation function (CCF) obtained with a dedicated A0 mask. The averaged stellar line has a measured projected velocity broadening of $v \sin i_{*}=112.6$~\kms. The vertical dashed line represent the systemic velocity (-17.74 \kms).}
\label{master_spectra}
\end{figure*}

We observed two transits of the ultra-hot Jupiter KELT-9~b around its bright A0V host star with the HARPS-N spectrograph \citep[][]{Cosentino2012} mounted on the 3.58 m Telescopio Nazionale Galileo (TNG) telescope in the Roque de los Muchachos observatory (La Palma, Spain)\footnote{Our observational strategy of recording several transits - to assess the reproducibility of the detected spectroscopic signatures - was made possible thanks to a Director's Discretionary Time proposal asked for in summer 2017 (A35DDT4) and a standard OPTICON proposal in the general frame of our SPADES survey (2018A/038, PI: Ehrenreich).}.

We dedicated a full night of continuous observations to each transit of KELT-9~b on 31 July 2017 and 20 July 2018 (see Table~\ref{tab:log}). One fiber recorded the target flux (fiber A) and the other monitored the sky (fiber B). In total, we registered 95 spectra, each with an exposure time of 600~s. The time series cover each transit in full and cover baselines of $\sim$2--3~h before and $\sim$1.5~h after each transit, used to build accurate out-of-transit master spectra. The typical signal-to-noise ratio (S/N) calculated in the continuum around 400 and 600 nm is between 27--131 and 70--190, respectively (no spectra were discarded in the analysis; however, the spectra collected on the second night are affected by a failure in the ADC, making their blue part much noisier than in the first night). Finally, the \texttt{CHESS.KING} module (KINematic and Geometric properties of the system) handles all the system parameters (see Table~\ref{tab:system}) needed during the data analysis.

We identified 44 spectra (22 in each night) fully in transit (i.e., entirely exposed between the first and fourth contacts) and we considered the 51 remaining spectra out of transit.
The HARPS-N raw frames are automatically reduced with the HARPS-N Data Reduction Software (DRS version 3.5). The DRS pipeline consists of an order-by-order optimal extraction of the two-dimensional echelle spectra. Then, for each exposure, the daily calibration set is used to flat-field, blaze correct, and wavelength-calibrate all the 69 spectral orders. Finally, the orders are merged into a one-dimensional spectrum and resampled  (with flux conservation) between 387~nm and 690~nm with a 0.001~nm wavelength step (the wavelengths are given in the air and the reference frame is the solar system barycenter one, see Fig.~\ref{master_spectra}). Our main analysis is conducted on the final HARPS-N merged spectra that have a spectral resolution of $\lambda/\Delta\lambda = 115,000$ or 2.7~\kms. In the present study, we took advantage of the cross-correlation functions (CCFs) that are computed with the DRS. The CCFs are constructed by correlating each spectral order with a weighted stellar mask \citep[a line list containing the main lines of the spectral type of interest,][]{Pepe2002}. Then, the data are further prepared with the \texttt{CHESS.ROOK} module (spectRoscopic data tOOl Kit). For the KELT-9 observations, we use a customized A0 mask \citep[as in][]{Anderson2018b} containing about 600 lines of typically \ion{Fe}{i}, \ion{Fe}{ii}, and other metal species. The A0 mask is sensitive to both the stellar and planetary spectra. Since the stellar spectrum is rotationally broadened, we computed the CCFs over a 700~\kms\ window around the systemic velocity and with an oversampling step of 0.082~\kms\ (a tenth of the pixel size). Then, we conducted the CCF analysis  on the sum of the CCF of all the 69 orders\footnote{we discarded (by visual inspection) the order number 1, 2, 50--58, 61, 64--66, 68 and 69 because they contain no or very few stellar lines, or because they are strongly affected by interstellar medium (ISM) or telluric line absorptions} (Fig.~\ref{master_spectra}). The use of a custom A0 mask boost the stellar CCFs contrast by a factor of $\sim$10 in comparison to the standard DRS reduction, allowing us to have a precision similar to the independent reduction presented in \citet{Borsa2019}.

The correction of the telluric absorption lines in each spectrum is the purpose of the \texttt{CHESS.KNIGHT} module (Knowledge of the NIGHtly Telluric lines). In this study, we used the \texttt{molecfit} software, version 1.5.7, provided by ESO \citep{Smette2015,Kausch2015}. Following the method presented in \citet{Allart2017} and applied in \citet{Hoeijmakers2018,Hoeijmakers2019,Seidel2019}, synthetic telluric spectra were computed with the real-time atmospheric profiles measured above the TNG/HARPS-N location and then fitted to each observed spectrum prior to dividing them. The correction of the changing telluric lines is measured to be at the noise level, especially for the $(4\nu + \delta)$ water band in the \Ha\ region. We note that the \Hb\ to \Hz\ regions are not affected by telluric lines.

%%%%%%%%%%%%%%%%%%%%%%%%%%%%%%%%%%%%%%%%%%%%%%%%%%%%%%%%%%%%%%%
\section{A comprehensive analysis of spectroscopic transit data}\label{Sec_newTS}
%%%%%%%%%%%%%%%%%%%%%%%%%%%%%%%%%%%%%%%%%%%%%%%%%%%%%%%%%%%%%%%

\begin{figure*}[t!]
\centering
\includegraphics[width=0.97\textwidth]{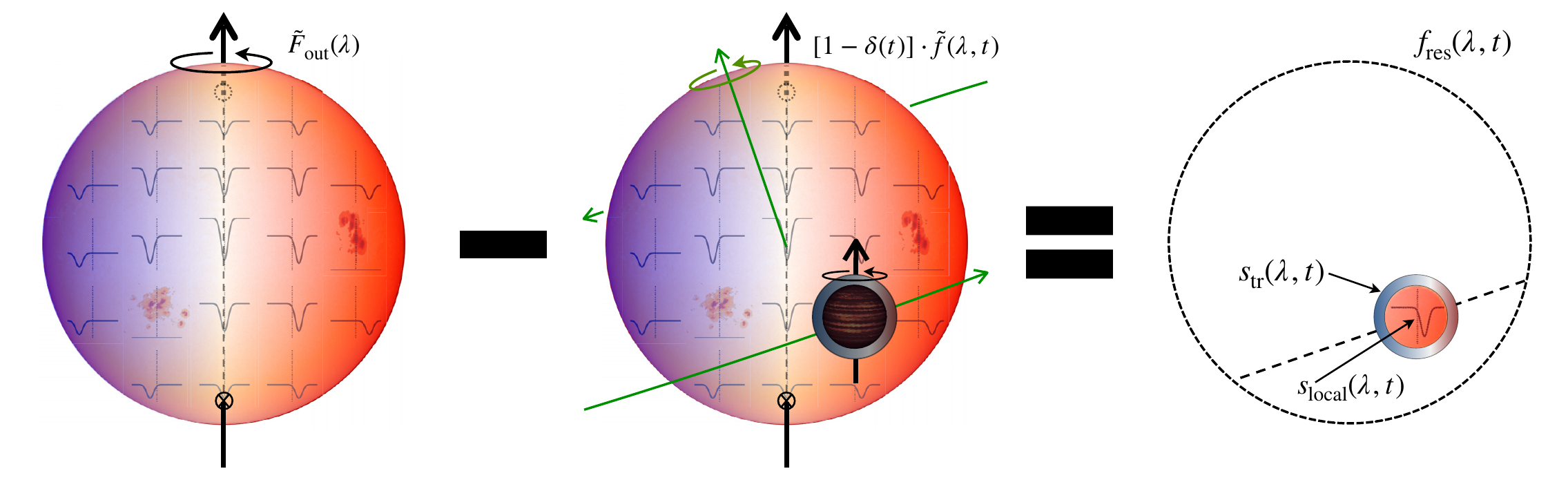}
\caption{Schematic of a spectroscopic transit observation of an exoplanetary system. The difference between an out-of-transit spectrum (left) and an in-transit spectrum (middle) gives us two essential quantities: the exoplanetary atmosphere transmission spectrum and the local stellar spectrum of the surface that is hidden behind the exoplanet (right). The temporal evolution of the stellar local spectrum will mostly depend on the projected stellar rotation (represented by the black arrow going through the star from pole to pole), the planetary orbit parameters (represented with the green arrows) and particularly the projected spin-orbit alignment angle. The temporal evolution of the exoplanet atmosphere spectrum will mostly depend on the atmosphere size, the planetary orbital projected velocity, and the local stellar spectrum that is transmitted through. Hence, it is necessary to separate and to measure accurately each component to retrieve the unknown parameters. The present figure is partially inspired by figures in \citet{Dravins2014,Dravins2017a,Dumusque2014,Cegla2016b,Bourrier2016b}.}
\label{Cartoon}
\end{figure*}

Spectroscopic transit observations (at high-resolution) have been widely used to characterize exoplanetary systems. On the one hand, they allow us to measure exoplanetary spin-orbit alignment, via the Rossiter-McLaughlin (RM) effect, with classical velocimetric measurements or with the tomographic and reloaded methods \citep{Queloz2000,Collier2010,Cegla2016b}. On the other hand, the same data also allow us to measure exoplanetary transmission spectra \citep{Wyttenbach2015}. These two signals can be entangled between each other or with any non-homogeneous stellar surface effects, such as the center-to-limb variation (CLV) of spectral lines, or stellar activity \citep{Czesla2015,Barnes2016,Cauley2017a}. The method presented here aims to separate and to accurately measure the stellar and planetary spectra (see Fig.~\ref{Cartoon}). Even though we are only taking some major effects into account, our method can be, in principle, adapted to correct for any stellar feature. An upcoming paper by Bourrier et al. will establish a rigorous formalism and present an accompanying pipeline \citep[but see also][who already used the same method]{Bourrier2020a,Ehrenreich2020}. Our method is implemented as in on our previous work \citep{Wyttenbach2015,Wyttenbach2017}, using the \texttt{CHESS} modules \texttt{BISHOP} (Basic Insights on the Stellar pHotosphere Occulted by the Planet) and \texttt{QUEEN} (QUantitative Extraction of Exoplanet atmospheres in transmissioN), and applied to the KELT-9 HARPS-N data.

%--------------------------------------------------------------------------------------------------------------------------------------------------------------------------
\subsection{On the extraction of a transmission spectrum}\label{Sec_spec_method}
%--------------------------------------------------------------------------------------------------------------------------------------------------------------------------

Because of the fluctuations of the Earth's atmospheric transmission, ground-based transit spectroscopy observations at high-resolution have to be self-normalized. The stellar spectra $f(\lambda,t)$ are divided by their respective values in a reference wavelength band $\langle\lambda_{\mathrm{ref}}\rangle$ chosen in the continuum near the region of interest to get the normalized spectra: $\tilde{f}(\lambda, t)\!=\!f(\lambda,t)/f(\langle\lambda_\mathrm{ref}\rangle,t)$. The master spectra in- and out-of-transit $F_\mathrm{in}(\lambda)\!=\!\sum_{t \in \mathrm{in}}f(\lambda,t)$ and $F_\mathrm{out}(\lambda)\!=\!\sum_{t \in \mathrm{out}}f(\lambda,t)$ are self-normalized in the same manner, giving $\tilde{F}_\mathrm{in}(\lambda)$ and $\tilde{F}_\mathrm{out}(\lambda)$. We now base our approach from the schematic shown in Fig.~\ref{Cartoon} and will proceed step by step.

We first treat an ideal case where there is no exoplanetary atmosphere, and we only consider the stellar rotation and the limb-darkening (and ignore other stellar effects). We are interested in extracting the Rossiter-McLaughlin effect. A possibility is to use the reloaded method \citep{Cegla2016b,Bourrier2016b,Bourrier2018a}:

\begin{equation}\label{eq:residuals}
f_{\mathrm{res}}(\lambda,t) = \tilde{F}_\mathrm{out}(\lambda) - [1 - \delta(t)] \cdot \tilde{f}(\lambda,t),
\end{equation}

where $1-\delta(t)$ represents the value of the photometric light curve (white light curve), and $\delta(t)$ is the transit depth at time $t$. We compute the light curve with a \citet{Mandel2002} model using the \texttt{batman} code \citep{Kreidberg2015} and the system parameters from Table~\ref{tab:system}. While the observed light-curve contains errors, these latter have a negligible effect on the normalization of the spectroscopic observations \citep{Cegla2016b,Bourrier2016b}. In this ideal case, $f_{\mathrm{res}}(\lambda,t_\mathrm{in})$ represents the stellar flux from the planet-occulted photosphere regions, that is to say the local stellar spectrum $s_\mathrm{local}(\lambda,t)$ emitted from the region behind the planet at time $t$. The measurement and modeling of $f_{\mathrm{res}}(\lambda,t_\mathrm{in})$ over time during the transit allow us to determine the sky-projected spin-orbit angle \citep{Cegla2016b}. The Rossiter-McLaughlin measurements are mostly done with CCFs in the velocity domain (because of the boost in signal). However, we underline that this measurement is also possible in a single spectral line.

We continue by treating a second ideal case where an exoplanet with an atmosphere transits a star deprived of its rotation (and of any other stellar features, except the limb-darkening). Since the star is not rotating, we know that the shape of the local stellar spectra $s_\mathrm{local}(\lambda,t)$ is not time-dependent and is equal to the out-of-transit spectrum $\tilde{F}_\mathrm{out}(\lambda)$. In this case, the only difference between the shape of the in- and out-of-transit spectra is the transmission through the planet atmosphere, implying that the stellar flux is multiplied by the planet apparent size. Then, the formula proposed by \citet{Brown2001} and implemented by considering the change in radial velocity of the planet in \citet{Wyttenbach2015}, is applied in a straightforward manner:

\begin{equation}\label{eq:TransitSpectrum_old}
-\mathfrak{R}^{'}(\lambda) = 1 - \frac{1}{{\rm N_{in}}}\sum\limits_{t \in \mathrm{in}} \left. \frac{[1 - \delta(t)] \cdot \tilde{f}(\lambda,t)}{\tilde{F}_\mathrm{out}(\lambda)}\right|_{p}.
\end{equation}

The $-\mathfrak{R}^{'}(\lambda)$ denotes the normalized spectrum ratio (averaged over ${\rm N_{in}}$ in-transit spectra), or the averaged transmission spectrum, and is equal to $(R_p(\lambda)/R_{\star})^2$. The $|_{p}$ expresses the spectrum Doppler shift according to the planet radial velocity and places the observer in the planet rest frame. As for the Eq.~\ref{eq:residuals}, we also use the light curve normalization $1-\delta(t)$ to put the in-transit spectra to their correct flux levels compared to the out-of-transit spectra. This correction is also necessary for the planetary atmosphere since the absorption depends on the relative stellar flux that is passing through the atmosphere \citep{Pino2018a}. This means that atmospheric absorptions were slightly overestimated in previous works. The calculation of Eq.~\ref{eq:TransitSpectrum_old} remains precise enough for slowly rotating stars \citep{Wyttenbach2017}.

We then present a comprehensive case that considers the presence of the planet atmosphere and an inhomogeneous stellar surface. The local stellar spectra $s_\mathrm{local}(\lambda,t)$  vary as a function of the stellar rotation and other effects (center-limb variation, convective blueshift, pulsations, gravity darkening, active regions, etc.). In transit configuration, we note that the light source transmitted through the planetary atmosphere is emitted from the region behind the atmosphere limb annulus\footnote{We ignore any refraction in the exoplanetary atmosphere.}. We assume that the light source $s_\mathrm{local,atm}(\lambda,t)$ coming from this annulus (with a surface ratio $\Sigma_\mathrm{atm}$ which is several times $2{\rm R_P H}/R_\star^2$) is equal in shape to the local stellar spectrum $s_\mathrm{local}(\lambda,t)$ (that comes from the surface behind the planet), but differs in intensity from it by a factor $\Sigma_\mathrm{atm}/\delta$, meaning that $s_\mathrm{local,atm}/\Sigma_\mathrm{atm}=s_\mathrm{local}/\delta$. The light source $s_\mathrm{local,atm}$ is absorbed by the exoplanet atmosphere and constitute $s_\mathrm{local,atm}\cdot s_\mathrm{tr}/\Sigma_\mathrm{atm}$, $s_\mathrm{tr}(\lambda,t)$ being the transmission spectrum we search to measure\footnote{We define the planetary absorption $A$ to be $A=s_\mathrm{tr}/\Sigma_\mathrm{atm}$.}. Owing to the absorption by the exoplanet atmosphere, the residuals $f_{\mathrm{res}}(\lambda,t)$, computed with Eq.~\ref{eq:residuals}, are not equal to the local stellar spectrum $s_\mathrm{local}(\lambda,t)$, but contain the sum of $s_\mathrm{local}(\lambda,t)$ and of $s_\mathrm{tr}(\lambda,t)\cdot s_\mathrm{local,atm}/\Sigma_\mathrm{atm}$. Hence, the spectrum ratio,

\begin{equation} \label{eq:TransitSpectrum}
s_\mathrm{tr}(\lambda,t) = \delta(t)\cdot\frac{f_{\mathrm{res}}(\lambda,t) - s_\mathrm{local}(\lambda,t)}{s_\mathrm{local}(\lambda,t)}
\end{equation}

is equal to the surface of the atmosphere limb (alone) times its wavelength dependent absorption, meaning that $\delta(t) + s_\mathrm{tr}(\lambda,t)=(R_p(\lambda,t)/R_{\star}(\lambda))^2$ is the transmission spectrum. Taking into account that the shape of the out-of-transit flux $\tilde{F}_\mathrm{out}(\lambda)$ is relative to the continuum reference band $\langle\lambda_{\mathrm{ref}}\rangle$, we get the final transmission spectrum formula relative to $\langle\lambda_{\mathrm{ref}}\rangle$:

\begin{eqnarray}\label{eq:TransitSpectrum}
-\mathfrak{R}^{'}(\lambda) &=& \frac{1}{{\rm N_{in}}}\sum\limits_{t \in \mathrm{in}} \left. \frac{\delta(t)+s_\mathrm{tr}(\lambda,t)}{\tilde{F}_\mathrm{out}(\lambda)}\right|_{p} \nonumber \\
\ &=&  \frac{1}{{\rm N_{in}}}\sum\limits_{t \in \mathrm{in}} \delta(t) \cdot \left. \frac{\tilde{F}_\mathrm{out}(\lambda) - [1 - \delta(t)] \cdot \tilde{f}(\lambda,t)}{\tilde{F}_\mathrm{out}(\lambda) \cdot s_\mathrm{local}(\lambda,t)}\right|_{p}.
\end{eqnarray}

Placing the observer in the planet rest frame is the last performed operation before the sum. We note that by setting $s_\mathrm{local}(\lambda,t)=\delta(t)~\forall\lambda$, we can simplify Eq.~\ref{eq:TransitSpectrum} to get Eq.~\ref{eq:TransitSpectrum_old}.

We first acknowledge that the presence of the planet atmosphere, as anticipated by \citet{Snellen2004}, magnifies the RM effect since the amplitude of the RM effect is proportional to the transit depth \citep[see also][]{Dreizler2009}. Additionally, at high-resolution, the planetary orbital velocity plays a major role. First, the change in shape of a stellar line due to the planet atmosphere happens when the planetary radial velocity is smaller than the full line broadening (which is dominated by the stellar $v \sin i_{*}$). Second, the line shape anomaly due to the RM effect is also altered by the planetary atmosphere when the two signals overlap in terms of velocity. Hence, the presence of a planet atmosphere biases the RM effect, in a way that is not a simple amplification, but rather by creating new anomalies that depend on the planetary radial velocity and of the spin-orbit angle. Despite \citet{Snellen2004} and \citet{Czesla2012} analyzed specific stellar lines to find RM anomalies, these latter were not modeled with the effect of the planet orbital velocity. This effect has been precisely taken into account only recently by \citet{Borsa2019}, who called it the ``atmospheric Rossiter-McLaughlin'' effect. This effect is particularly important for ultra-hot Jupiters who share many spectral lines with their host stars \citep[see more details in][]{Bourrier2020a,Ehrenreich2020}.

Similarly to the RM effect being modified by the planet atmosphere, the planetary atmosphere absorption depends on the stellar flux that is passing through it. For example, when the local stellar and the atmospheric spectra overlap, the atmospheric absorption is lowered because less flux is coming from the star. This effect is automatically taken into account and corrected for in Eq.~\ref{eq:TransitSpectrum} thanks to the division by $s_\mathrm{local}(\lambda,t)$.

This method relies on the knowledge of the local stellar profile $s_\mathrm{local}(\lambda,t)$. One approach is to model $s_\mathrm{local}(\lambda,t)$ by using stellar spectrum model at different limb-darkening angles \citep[e.g.,][]{Yan2017,Yan2018,Casasayas-Barris2019}. Our approach is to empirically fit the observed local stellar profile (see Sec.~\ref{Sec_Puls}). We emphasize the caveat, particularly during the overlap, of the uncertainty whether the local stellar profile remains constant.

%--------------------------------------------------------------------------------------------------------------------------------------------------------------------------
\subsection{Analysis of the local stellar photospheric CCFs}\label{Sec_RM}
%--------------------------------------------------------------------------------------------------------------------------------------------------------------------------

\begin{figure}[t!]
\centering
\includegraphics[width=0.47\textwidth]{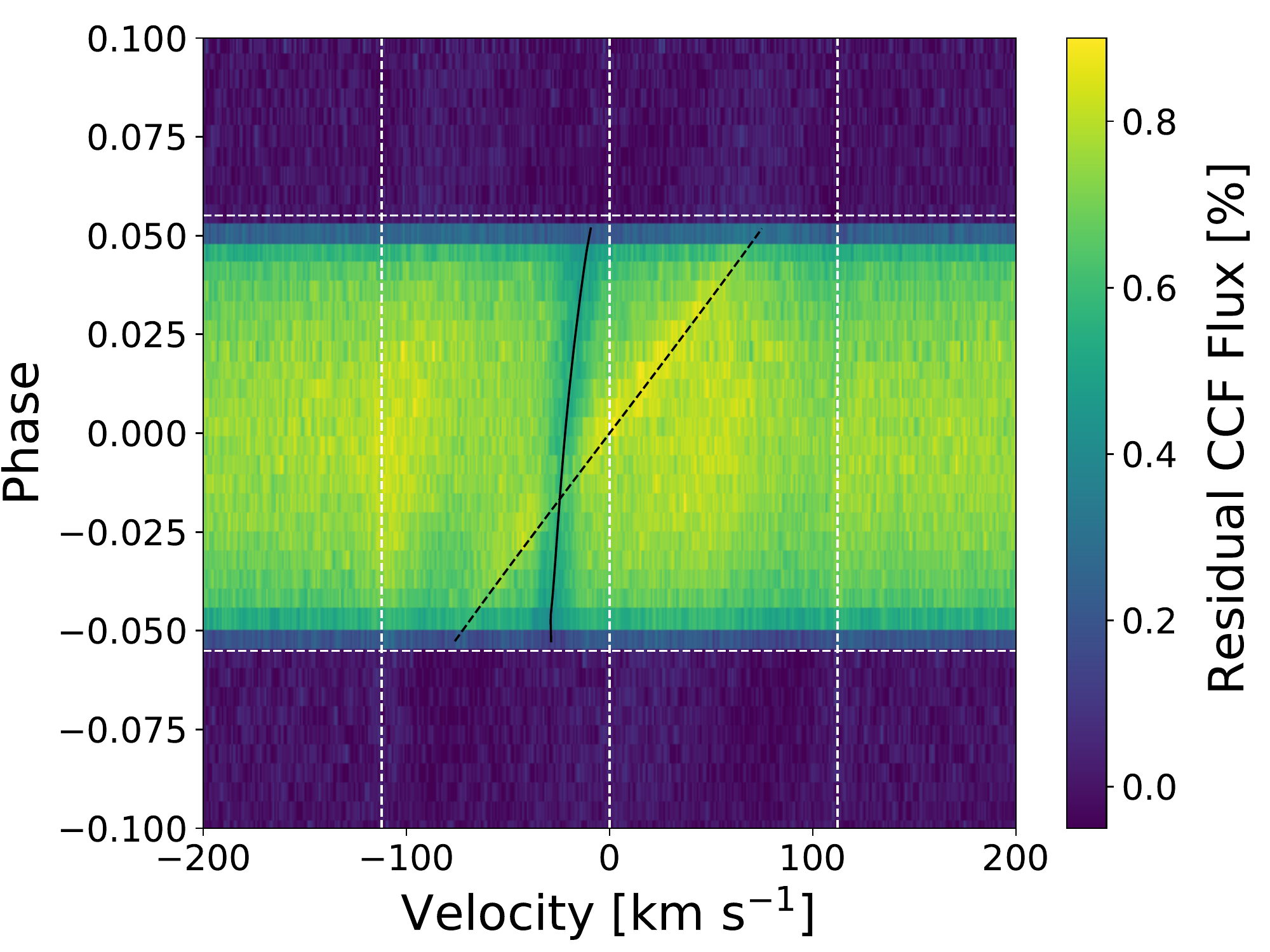}
\includegraphics[width=0.47\textwidth]{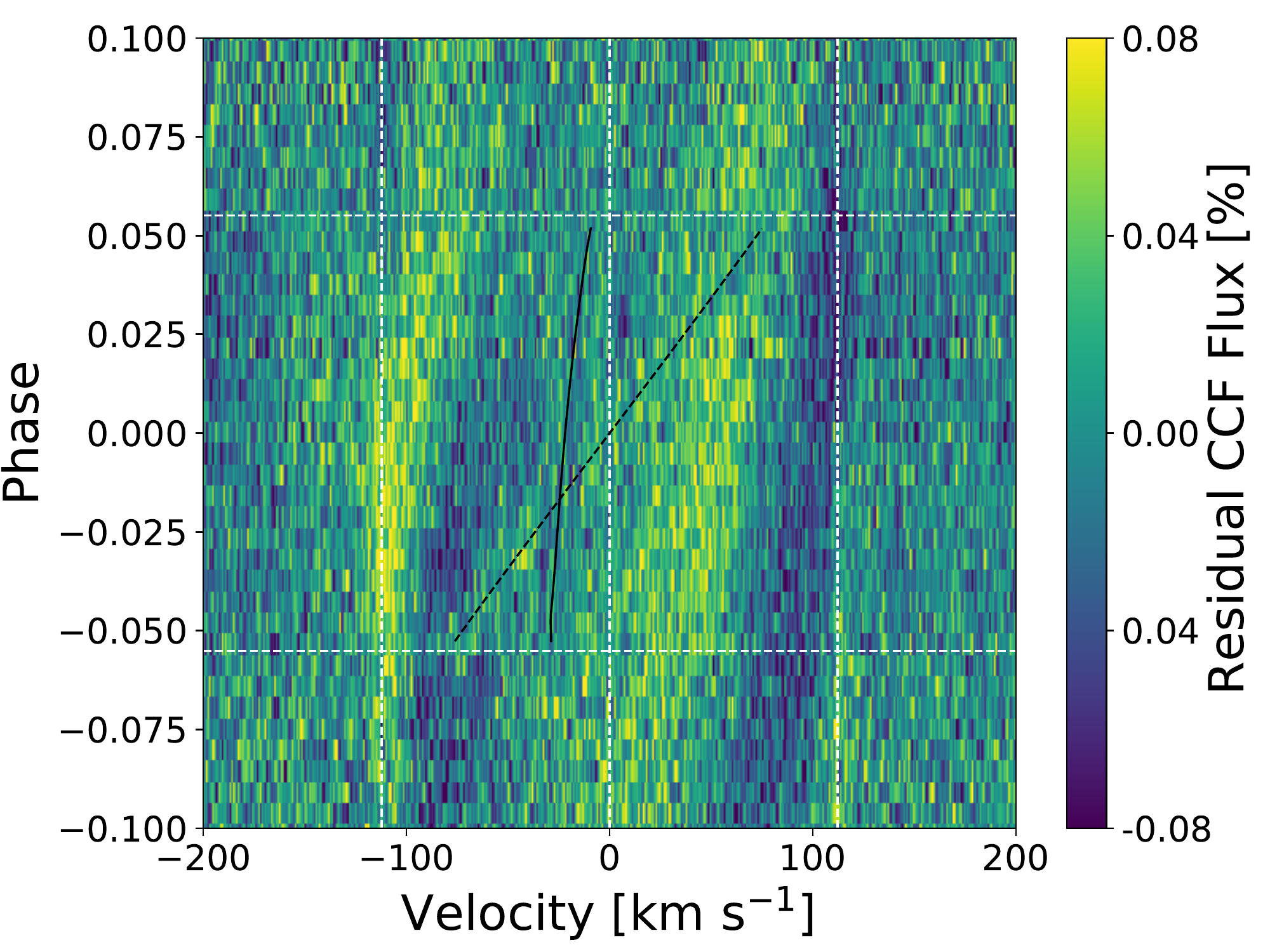}
\caption{Map of the residual CCF flux during the transit of KELT-9 b. The observations of two nights are binned by 2 in phase. The 1st and 4th contacts are traced with the horizontal dashed white lines; the 0~\kms\ and $\pm v \sin i_{*}$ are traced with vertical dashed white lines. Top panel: Local stellar spectrum tracing the spin-orbit misalignment through the Rossiter-McLaughlin effect. The solid black line shows our best fit. The exoplanet transmission CCF, that has a slanted shape following the orbital velocity (dashed black line), traces the metal detections by \citet{Hoeijmakers2018,Hoeijmakers2019}. Bottom panel: Residual CCF flux after the removal of the local stellar and of the planetary atmosphere spectra. The residuals are dominated by the stellar pulsations, whose strengths are 10$\times$ lower than the local stellar signal (see the color scale). The pulsations are due to pressure-mode in a $\delta$ Scuti-type star.}
\label{fig:CLB}
\end{figure}

The analysis of the stellar CCFs is conducted on the CCFs constructed with the aforementioned A0 mask \citep[similarly to][]{Anderson2018b,Borsa2019}. The CCF residuals are computed with Eq.~\ref{eq:residuals} in the velocity domain (see Fig.~\ref{fig:CLB}). They clearly show the local stellar CCFs (the CCF of the spectrum of the stellar surface that is hidden behind the exoplanet, see Sec.~\ref{Sec_RM_model}), the CCFs of some stellar pulsations (see Sec.~\ref{Sec_Puls}) and the planetary atmosphere CCFs trace. The planet atmospheric trace is an independent confirmation of the metal (mainly \ion{Fe}{i}, \ion{Fe}{ii}) detection of \citet{Hoeijmakers2018,Hoeijmakers2019} and \citet{Borsa2019}, and is not further discussed in this study. The main goal of this section is to focus on the pulsation signal, and then on the local stellar CCFs to have an updated value of the spin-orbit angle $\lambda$ and to use this value as a prior in our Balmer lines analysis.

%--------------------------------------------------------------------------------------------------------------------------------------------------------------------------
\subsubsection{Spectroscopic detection of stellar pulsation}\label{Sec_Puls}
%--------------------------------------------------------------------------------------------------------------------------------------------------------------------------

\begin{figure}[t!]
\centering
\includegraphics[width=0.47\textwidth]{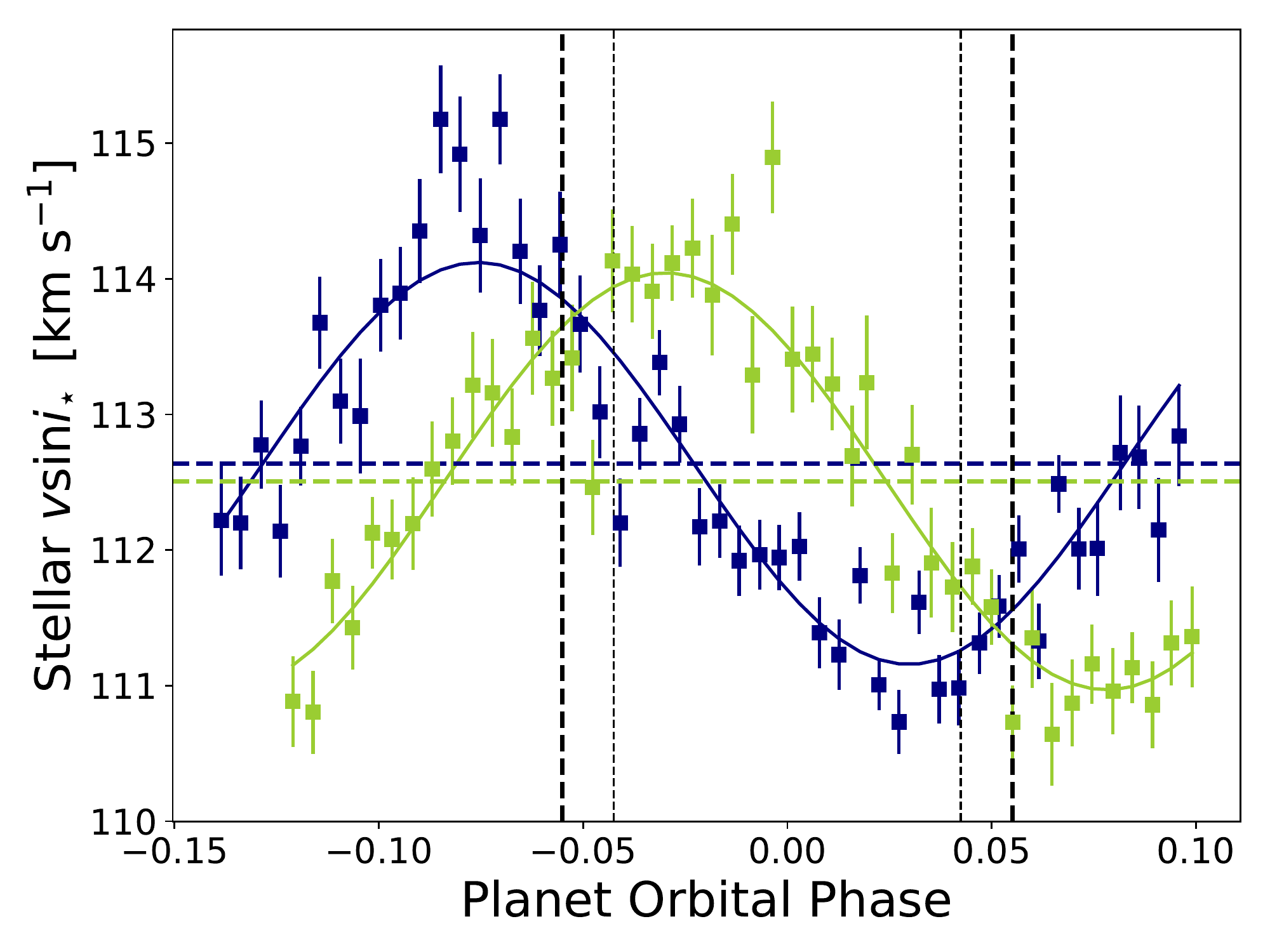}
\caption{Stellar pulsation signal seen from variation in the stellar $v \sin i_{*}$. The stellar CCF are cleaned from the local stellar and planetary atmosphere spectra, and fit with a stellar rotation profile. The pulsation period is ${P_{\rm puls}}=7.54\pm0.12$~h, and is consistent between the two nights (in dark blue and green, respectively) that are separated by one year. The averaged $v \sin i_{*}$ is reported for each night with the horizontal dashed lines, and has an averaged value of $v \sin i_{*}=112.60\pm0.04$~\kms.}
\label{fig:CLB_puls}
\end{figure}

Prior to discussing the reloaded RM study, we first mention the spectroscopic detection of stellar pulsation. As all the different signals can be disentangled, we study the CCFs cleaned from the local stellar and the planetary atmosphere CCFs, but still keep the pulsation signals (Fig.~\ref{fig:CLB}). A rotationally broadened model (including convolution with the instrumental profile and with a local stellar CCFs width) are fitted to each CCF \citep{Gray2005,Anderson2018b,Borsa2019,Dorval2019}. We measure an averaged $v \sin i_{*}$ value of $112.60\pm0.04$~\kms. We note that our CCFs (see Fig.~\ref{master_spectra}) and $v \sin i_{*}$ values are similar to those published in the results of \citet{Borsa2019}. The instrumental profile (2.7~\kms) with the local stellar CCFs (see Sec.~\ref{Sec_RM_model}) have an averaged broadening of ${\xi}=9.17\pm0.05$~\kms, making the total broadening $\sim$113~\kms. The stellar local profiles are fully described by Gaussians. Indeed, the local stellar profile have a small broadening, despite the high disk integrated rotational broadening ($\xi\ll v \sin i_{*}$). This is because of the high observing cadence, the small impact parameter, and the inclined orbit (see Sec.~\ref{Sec_RM_model}). We measure a sinusoidal oscillation in the time series of the stellar $v \sin i_{*}$ around the averaged value. The oscillation signal is not in phase with the planetary transit. However, this signal has a period $P=7.54\pm0.12$~h, in phase within the two nights separated by one year (Fig.~\ref{fig:CLB_puls}). This period corresponds to the stellar pulsation signal seen in the KELT-9 TESS light curve of ${P_{\rm puls}}=7.5868\pm0.0009$~h \citep{Wong2019}, and is compatible with pressure modes (p-mode) in a $\delta$ Scuti-type star. This is therefore an independent spectroscopic detection of the KELT-9 $\delta$ Scuti-type pulsation. The first and second CCF moments are clearly described by a single sine function with a period ${P_{\rm puls}}$ (the CCF radial velocities are the first CCF moment, and the $v \sin i_{*}$ are proportional to the second CCF moment), hinting to a sectoral mode oscillation \citep[the azimuthal order of the mode $m$ is equal to $\pm l$, the degree of the mode, see e.g.,][]{Balona1986,Aerts1992,Aerts2010}. A complete analysis of the pulsation mode is out of the scope of this paper, but a follow-up study could bring complementary information to the spin-orbit analysis, in particular, for the stellar inclination.

%--------------------------------------------------------------------------------------------------------------------------------------------------------------------------
\subsubsection{``Reloaded'' Rossiter-McLaughlin analysis}\label{Sec_RM_model}
%--------------------------------------------------------------------------------------------------------------------------------------------------------------------------

\begin{figure}[t!]
\centering
\includegraphics[width=0.47\textwidth]{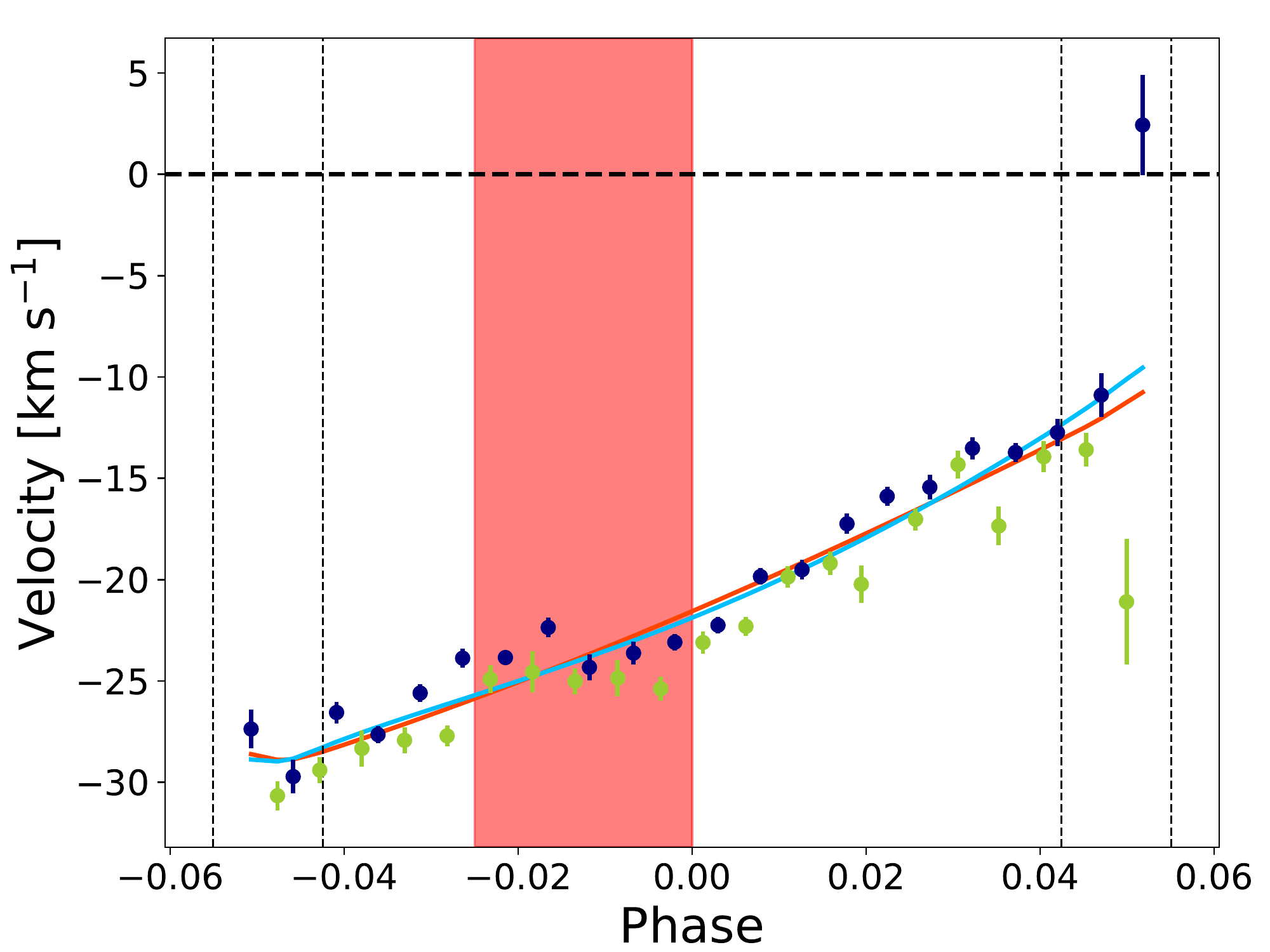}
\caption{Velocities of the local stellar surface hidden behind the planet during its transit. The observations of the two nights are shown in dark blue and green, respectively. The first to fourth contacts are traced with the vertical dashed lines. The overlap region between the local stellar and atmospheric signals (between phases -0.025 and 0) is shown in red and is not used for the fit. The spin-orbit misalignment is modeled with a solid body rotation (in orange), and with a differential rotation (in light blue). The differential rotation model is slightly preferred.}
\label{fig:CLB_vel}
\end{figure}

For the reloaded RM analysis, we treated the CCF residuals in a similar way to  that of recent studies by \citet{Bourrier2020a,Ehrenreich2020}. We first removed the signals from the stellar pulsations in the out-of-transit phases by fitting and removing as many Gaussians as there are of individual pulsation signals, allowing us to recompute a new out-of-transit master, from which we recompute again the CCF residuals. At this stage, we remove the stellar pulsation in-transit and the planet atmosphere signal with Gaussian fitting. By doing so, we are left with the local stellar CCF alone ($s_\mathrm{local}$; see e.g., \citet{Cegla2016b,Wyttenbach2017,Bourrier2020a} for more descriptions of the local stellar CCF), from which we measure each local velocity (Fig.~\ref{fig:CLB_vel}). We discard the velocities coming from the overlapping region with the atmospheric signal, because they are biased \citep{Borsa2019}. Local stellar CCFs are considered to be contaminated by the planet atmosphere if the local CCF full width at half maximum (FWHM) overlaps the planetary atmosphere CCF FWHM (10 data points between phase -0.025 and 0.000). We noticed a posteriori that discarding or keeping the overlapping region changes our results by about 1~$\sigma$. We also discard the outliers points at the egress, for which the CCFs (and velocities) are not detected following the S/N criteria of \citet{Cegla2016b} and \citet{Bourrier2018a}.

We model the residual CCF velocities with the formalism of \citet{Cegla2016b}, that computes the brightness-weighted average rotational velocity behind the planet during each exposure. The model parameters are the spin-orbit angle, and the stellar velocity field. The latter could be described by a rigid rotation (with $v \sin i_{*}$ as a parameter) or a solar-like differential rotation law, which could be expected for A-type star \citep{Reiners2004,Balona2016}. In this case, we have the three parameters $v_{\rm eq}$, $i_{*}$, and $\alpha$ which describe a solar-like differential rotation law \citep{Cegla2016b}. We use \texttt{emcee} \citep{Foreman-Mackey2013} to explore the parameter space and to find the best models and their parameter confidence intervals (CI). For each of the two nights and the combination of the two nights, we use 200 walkers for 1500 steps and a burn-in size of 500 steps. We use the Bayesian information criterion (BIC) to compare between models. For the solid rotation case (BIC=151.5), we found $v \sin i_{*}=122.5\pm0.6$~\kms, and $\lambda=\ang{-85.65}\pm\ang{0.09}$ (see Fig.~\ref{fig:RM_SB}). The $v \sin i_{*}$ differs between the two nights and is also bigger than our spectral fit, and the one of \citet{Borsa2019}. Despite this, the spin-orbit angle is consistent with \citet{Gaudi2017} and \citet{Borsa2019}. The differential rotation case (BIC=147.4) has a $\Delta{\rm BIC}=4.1$ in its favor, which is interpreted as ``positive evidence'' but not a confirmed detection ($\sigma$$\sim$2.2). In this scenario, we found $\alpha=-0.10^{+0.03}_{-0.04}$, $v_{\rm eq}=195^{+37}_{-23}$~\kms, $i_{*}=\ang{143.2}^{+\ang{6.3}}_{-\ang{6.0}}$, and $\lambda=\ang{-85.01}\pm\ang{0.23}$ (best solution, see Fig.~\ref{fig:RM_DR_neg}). We note that $\lambda$ is still compatible (at 3~$\sigma$) with the rigid rotation case, and that the corresponding $v \sin i_{*}=116.9\pm1.8$~\kms\ is more consistent with our spectral fit. Another differential solution, with a $\Delta{\rm BIC}=3.8$, was found with $\alpha=0.09\pm0.03$, $v_{\rm eq}=155\pm18$~\kms, $i_{*}=\ang{55.9}^{+\ang{11.6}}_{-\ang{7.2}}$, and $\lambda=\ang{-85.10}\pm\ang{0.25}$, with a very similar local stellar trace to the previous solution, but with a higher $v \sin i_{*}=127.7^{+2.7}_{-2.4}$~\kms (see Fig.~\ref{fig:RM_DR_pos}). Our two differential rotation solutions give a true spin-orbit angle of $\psi=\ang{89.6}\pm0.7$ and $\psi=\ang{84.2}^{+\ang{0.4}}_{-\ang{0.3}}$, respectively. The best fit, shown in Figs.~\ref{fig:CLB} and~\ref{fig:CLB_vel},  is later used for our Balmer lines extraction.

The differential rotation solutions give equatorial velocities that are typical of A-type stars \citep{Zorec2012}; furthermore an A-type star is likely to show differential rotation \citep{Zorec2017}. The different solutions have a solar-type differential rotation ($\alpha=0.08$), and an ``anti-solar''-type ($\alpha=-0.1$), respectively, meaning that in the latter case, that the high latitudes are rotating slightly faster than the equator. The equatorial velocities found imply that the stellar rotation periods are ${\rm P_{rot}}=0.7\pm0.1$~day ($\sim$17.1~h) and ${\rm P_{rot}}=0.56\pm0.09$~day ($\sim$13.6~h), respectively. We note that these periods do not correspond to the pulsation period of $\sim$7.6~h. This means that the star is turning at $\sim$33--56\% of its theoretical critical Keplerian break-up velocity ($v_{\rm crit}\sim416$~\kms). With such an equatorial velocity, the angular rotation parameter $\omega=\Omega_{\rm rot}/\Omega_{\rm crit}$ is about 0.45--0.75 \citep{Maeder2009}. This implies that gravity darkening effects become significant and measurable, for example by TESS or high quality photometry \citep[as already done by][]{Wong2019,Barnes2011,Ahlers2015,Ahlers2020}. For example, with $\omega=0.45$--0.75, the star becomes oblate and the equatorial radius becomes up to $\sim4$--12\% bigger than the polar radius \citep{Georgy2014}. In a gravity darkened star, the polar regions are hotter and more luminous than the equatorial regions. Because of the stellar inclination measured in the differential rotation scenario, we are mainly measuring light from the polar region of the star. Hence, at $\omega=0.45$--0.75, the stellar luminosity is overestimated up to $\sim5$--15\%, and the effective temperature up to $\sim1$--3\%, depending on the stellar inclination toward the observer \citep{Georgy2014}. This may be the source of the differences in retrieved parameters between \citet{Gaudi2017} and \citet{Borsa2019}. Thus, a follow-up study with high-resolution spectroscopy and precise photometry, which take care of gravity darkening effects, must be undertaken, and will allow us to refine our solution and to choose between different degenerate geometries \citep[e.g.,][]{Zhou2019,Dorval2019,Ahlers2020}. In particular, the gravity darkening provides a third possibility to measure the stellar inclination alongside to the Rossiter-McLaughlin effect and the stellar pulsation mode analysis. Finally, because of the polar orbit, the small impact parameter and the stellar inclination, the planet may be transiting near the pole during the transit \citep{Wong2019}. Such a transit configuration is favorable to measure star-planet interaction because of anisotropic stellar flux and winds \citep[``gravity-darkened seasons'',][]{Ahlers2020}.

%--------------------------------------------------------------------------------------------------------------------------------------------------------------------------
\subsection{The transmission spectrum of KELT-9~b in the Balmer lines regions}\label{Sec_TS}
%--------------------------------------------------------------------------------------------------------------------------------------------------------------------------

To extract the planetary atmosphere information from the spectra, we follow the procedure that leads to Eq.~\ref{eq:TransitSpectrum}. We assume the system parameters from Table~\ref{tab:system}. In particular, for each Balmer line we use the same white light curve from \citet{Gaudi2017} to normalize the spectra because it is the only one available\footnote{This has little impact since the expected relative change of the line contrasts would be less than $0.5\%$ \citep[from the prediction of Fig. 14 of][]{Yan2019}, that is smaller than our individual error bars.}. To correct the transmission spectra from the local stellar photospheric signal, we must know $s_\mathrm{local}(\lambda,t)$ for every Balmer line. We assume that the local stellar signal in the Balmer lines follows the CCF one in terms of velocity. For each Balmer line, and for each phase, we fit a Gaussian to the local stellar spectrum, with the velocity centers following our best reloaded RM model from Sec.~\ref{Sec_RM}. The use of the differential rotation model instead of the rigid rotation model does not change our final line extraction. Indeed, the two models are similar and the velocity errors from the CCFs are much smaller than the one present in a single Balmer line. Thus, we can fix the velocities to the best reloaded RM model in our fit. The Gaussian offset values follow the light curve $\delta(t)$. We therefore only fit for the contrast and the FWHM values. In a second step, we averaged all the contrast and FWHM values, and fitted again a Gaussian (for each line, for each phase) with starting positions coming from the averaged values, and by not allowing us the fit to exceed $\pm1\sigma$ value of the averaged contrast and FWHM. Whenever the local stellar spectrum signal is too weak for the Gaussian fit to converge (or to be significant), we assume the averaged contrast and FWHM values. We note that the stellar pulsations, despite being expected in the Balmer lines, are too weak to be detectable given the S/N of a single line. Thus, the pulsation contaminations in the Balmer lines are at the noise level. Also, we removed low-frequency trends in the residual continuum by fitting polynomials \citep{Seidel2019}. The knowledge of $s_\mathrm{local}(\lambda,t)$, and the $K_p$ from \citet{Hoeijmakers2019} allows us to compute the weighted averaged transmission spectra with Eq.~\ref{eq:TransitSpectrum}\footnote{In practice, Eq.~\ref{eq:TransitSpectrum} is used with a weighted mean that take the changing data quality into account. The weights $w$ at time $t$ are the inverse of the squared flux uncertainties, $w(t)\!=\!1/\sigma(t)^2$.}. We discard the spectra where the atmospheric signal overlaps with the local stellar signal (-0.027$\leq$Phase$\leq$0.0015), because we cannot fit $s_\mathrm{local}(\lambda,t)$ with a gaussian for these phases. Figure~\ref{fig:RM_TS_Ha} shows a representative example of the effect of the local stellar photospheric signal correction (RM effect correction). We measure the RM effect to have the same relative amplitude for the different Balmer lines and to be corrected down to the noise level. We compute the errors on the absorption lines by considering the systematic noise, empirically measured by the standard deviation into the continuum. We also consider the stellar line contrasts that decrease the S/N in the absorption region (see Table~\ref{tab:log}). This is particularly relevant for the blue part of the detector. Each line center takes the systemic velocity into account (e.g., 6562.408~$\AA$ for \Ha); the reference passbands for the normalization of every spectrum are centered 6~$\AA$ on each side of the line center, and are 6~$\AA$ wide (e.g., [6553.408:6559.408]$\cup$[6565.408:6571.408] for \Ha). For the different Balmer lines, our main results are:

\begin{figure*}[t!]
\centering
\includegraphics[width=0.47\textwidth]{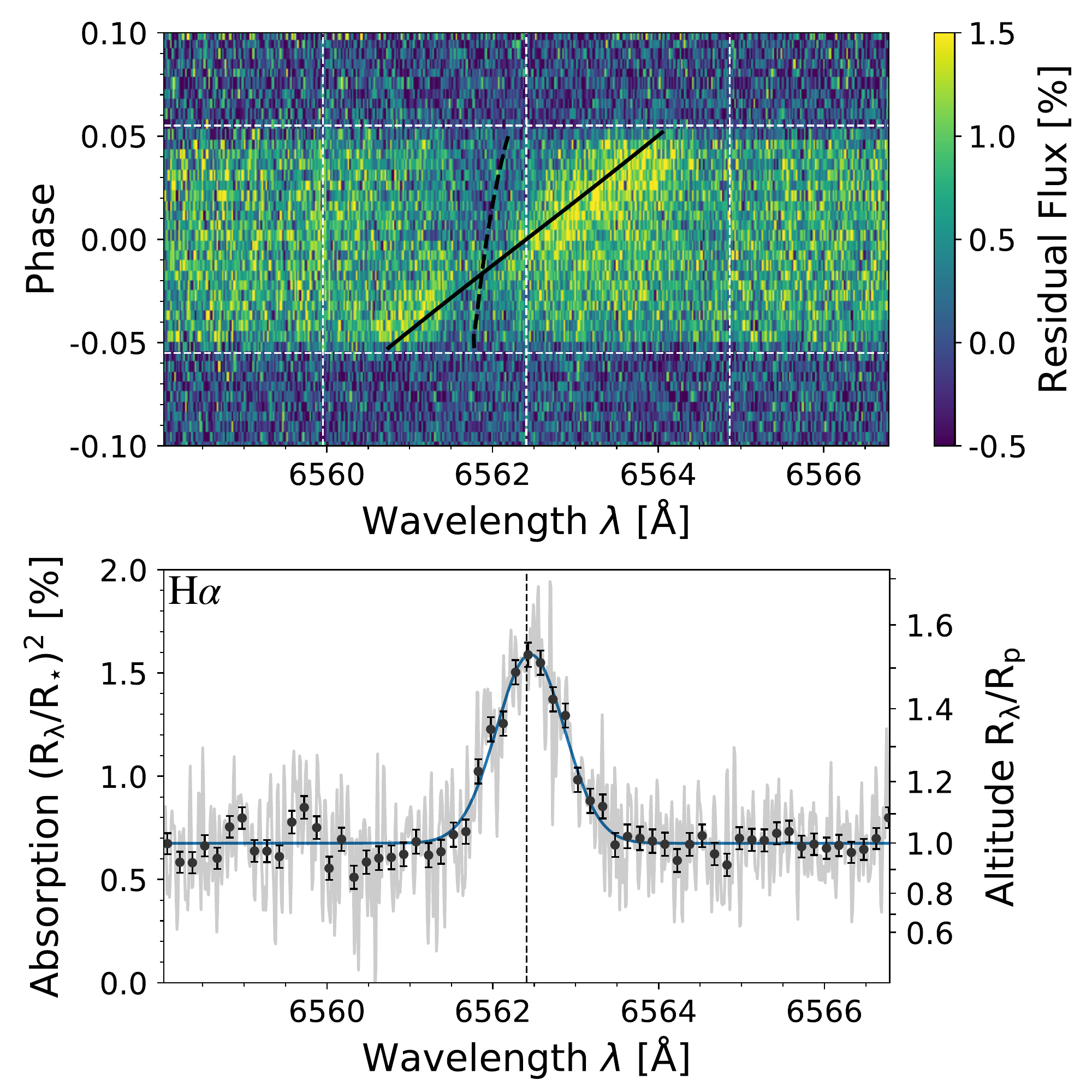}
\includegraphics[width=0.47\textwidth]{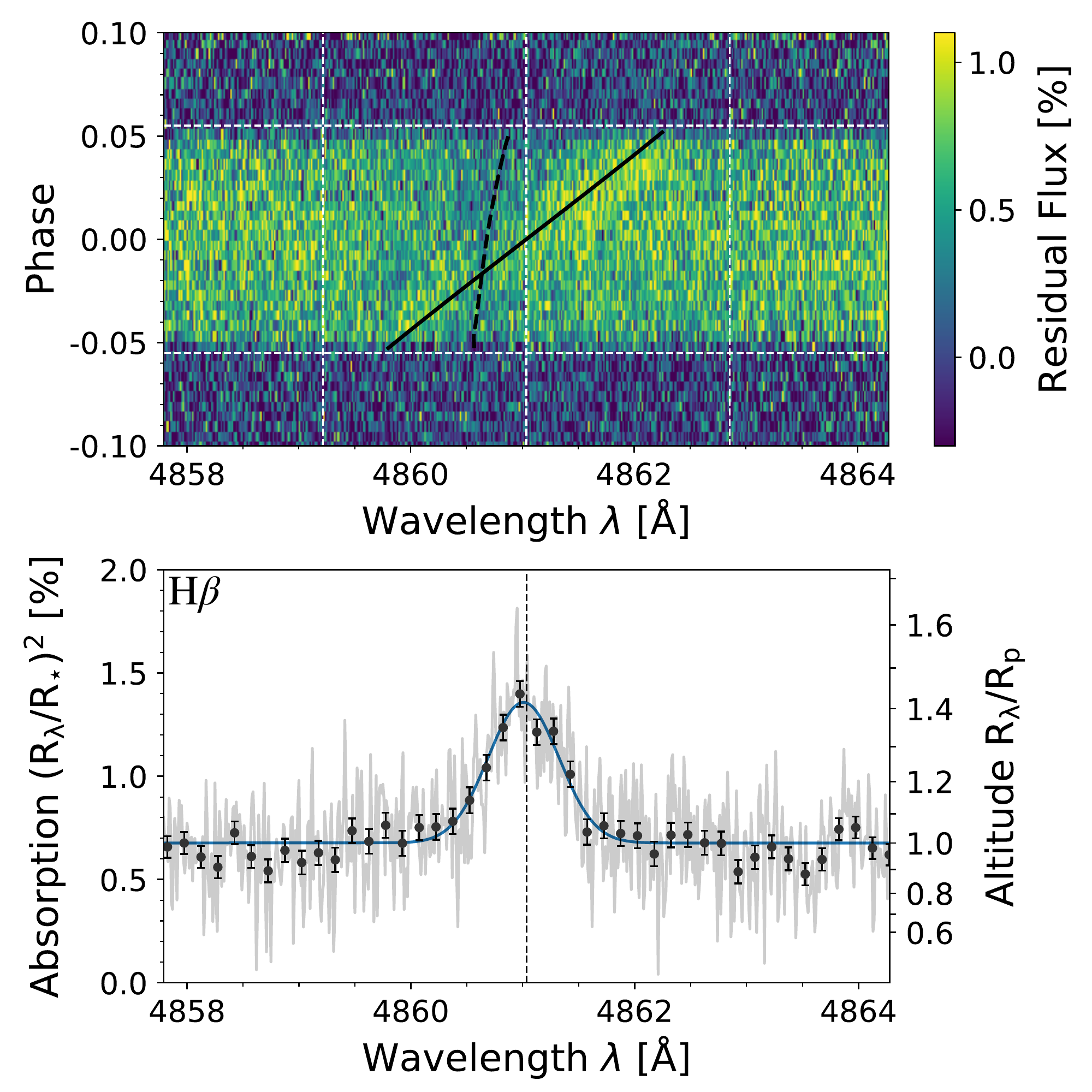}
\caption{Left: Detection of the planetary \Ha\ line (significance of 24.3~$\sigma$). Top left panel: Map of the residual spectra $\pm200$~\kms\ around the \Ha\ line in the stellar rest frame (Eq.~\ref{eq:residuals}). The observations of two nights are binned by 2 in phase. The planetary atmosphere \Ha\ absorption follows the orbital velocity (solid black line). The local \Ha\ stellar line follows the reloaded RM model (dashed black line). The first and fourth contacts are traced with the horizontal dashed white lines, the 0~\kms and $\pm v \sin i_{*}$ are traced with vertical dashed white lines. Bottom left panel: The weighted averaged \Ha\ transmission spectrum (corrected from the local stellar photospheric signal, Eq.~\ref{eq:TransitSpectrum}) in the planet rest frame (light gray), also binned by $15\times$ in black circles (the continuum is set to the white light curve transit depth $\delta=0.00677$). The vertical black dashed line shows the line center taking the systemic velocity into account. We show the Gaussian fit (blue) with contrast of $0.91\pm0.04\%$ and FWHM of $44.3\pm1.8$~\kms. Right: Same as for \Ha,\, but for the \Hb\ line (detection significance of 16.1~$\sigma$, see Table~\ref{tab:line_fit}).}
\label{fig:Ha_Hb_lines}
\end{figure*}

\begin{figure*}[t!]
\centering
\includegraphics[width=0.47\textwidth]{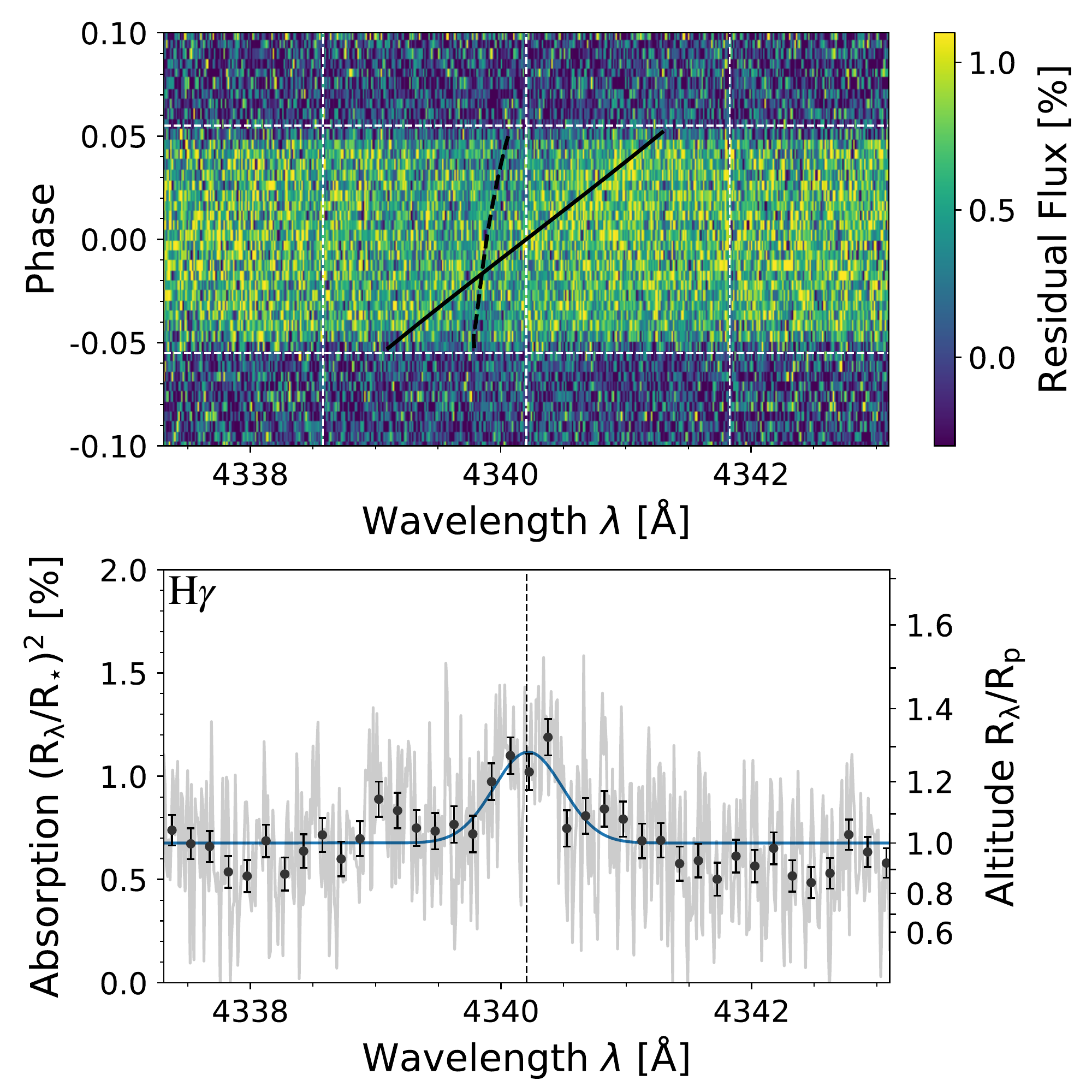}
\includegraphics[width=0.47\textwidth]{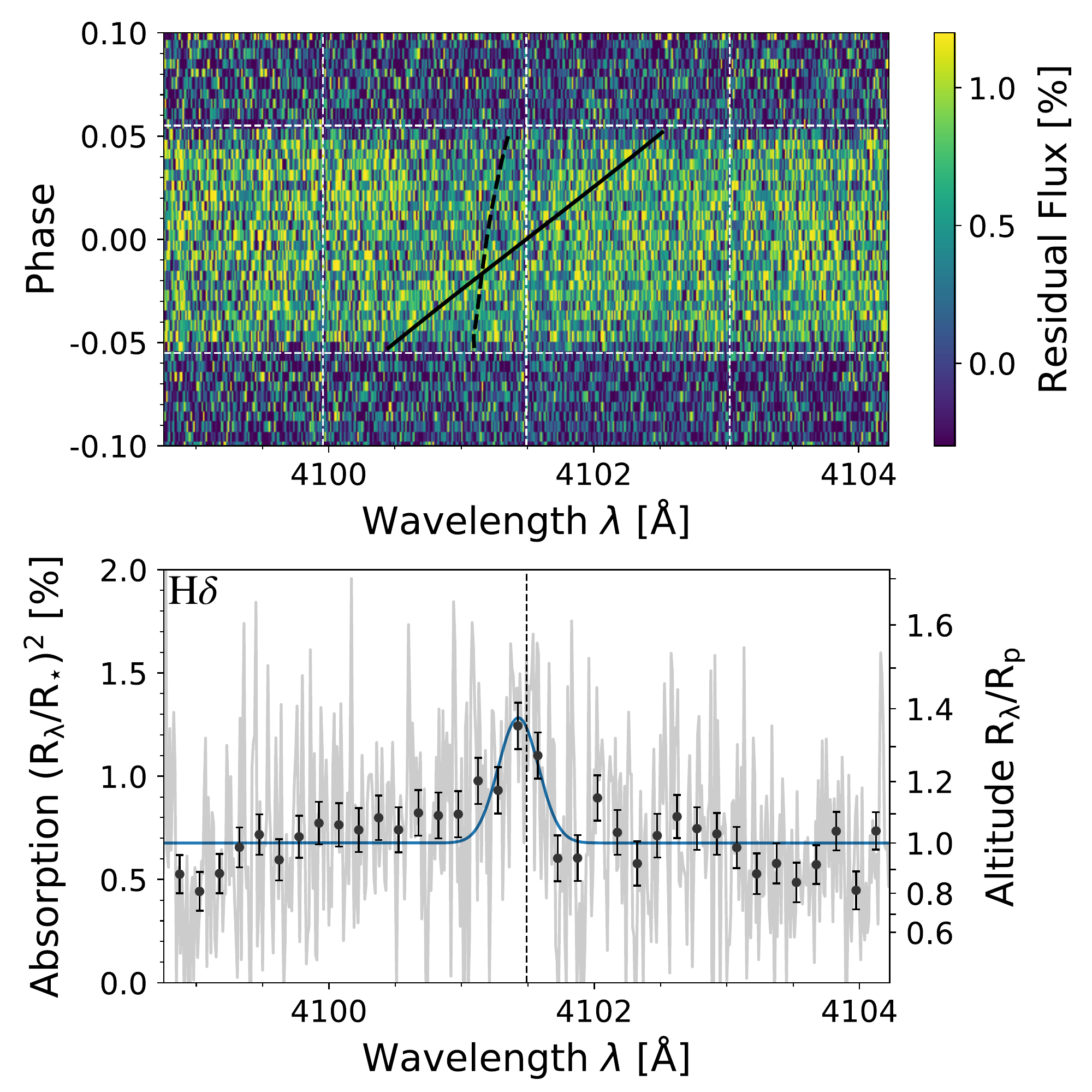}
\caption{Same as for Fig.~\ref{fig:Ha_Hb_lines}, but for the \Hg\ line (left) and the \Hd\ line (right). The \Hg\ and \Hd\ lines are detected with a significance of 7.5~$\sigma$ and 4.6~$\sigma$, respectively (see Table~\ref{tab:line_fit}).}
\label{fig:Hg_Hd_lines}
\end{figure*}

\begin{figure*}[t!]
\centering
\includegraphics[width=0.47\textwidth]{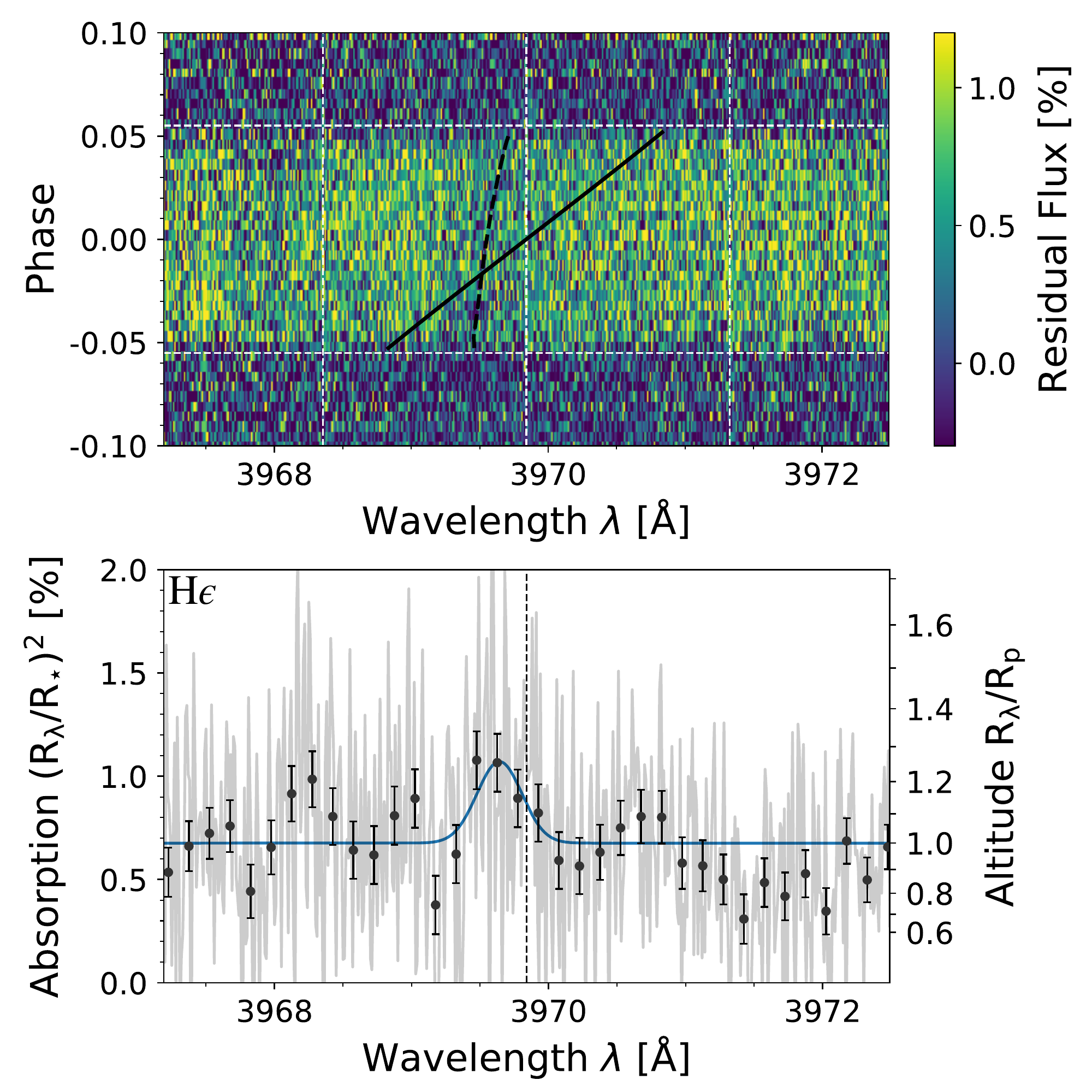}
\includegraphics[width=0.47\textwidth]{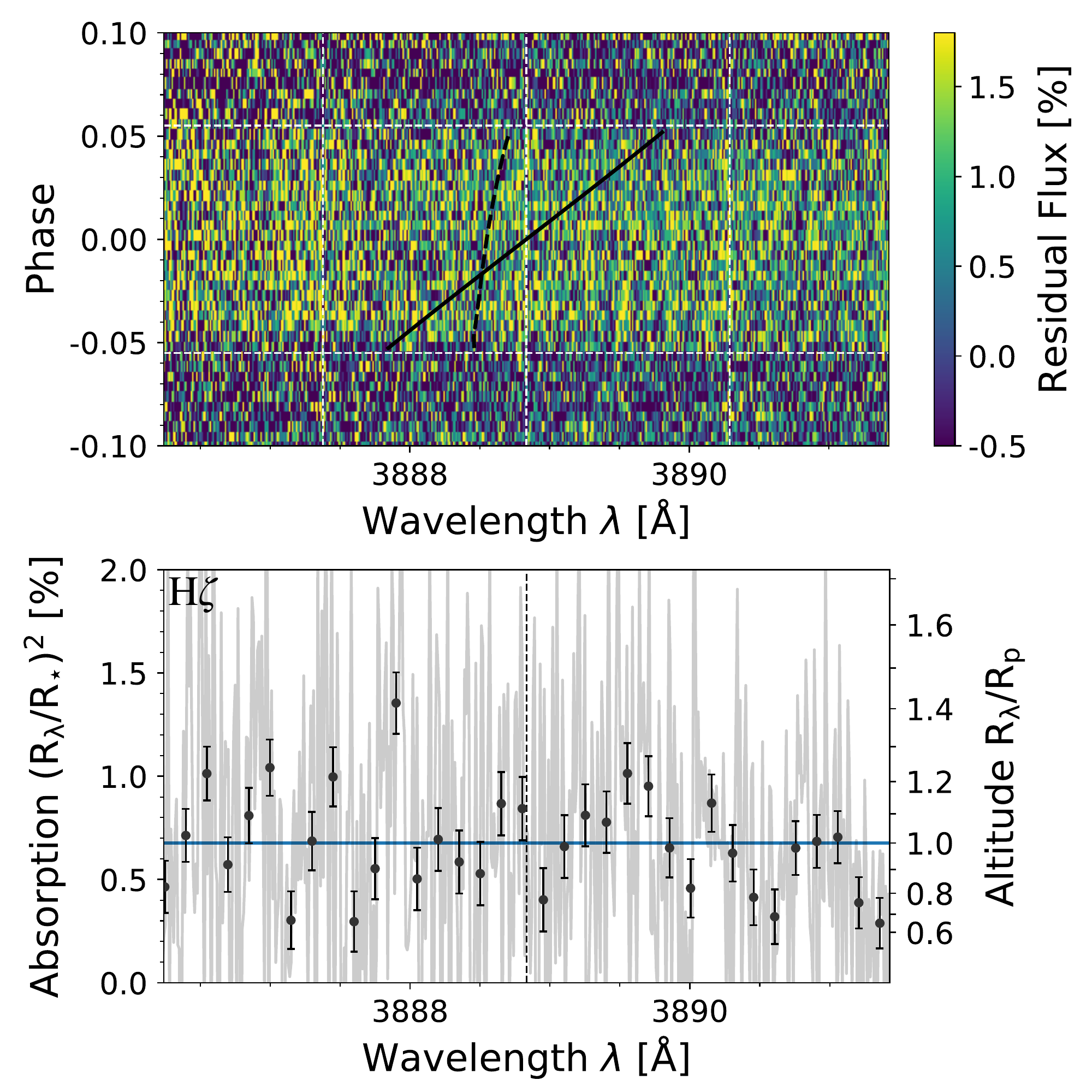}
\caption{Same as for Fig.~\ref{fig:Ha_Hb_lines}, but for the \He\ line (left), and the \Hz\ line (right). The \He\ and \Hz\ lines are not statistically detected (see Table~\ref{tab:line_fit}).}
\label{fig:He_Hz_lines}
\end{figure*}

\begin{itemize}
\item \Ha: The \Ha\ line is detected with a high confidence of 24.3~$\sigma$, which makes the planetary trail visible for each phase (Fig.\ref{fig:Ha_Hb_lines}). Fig.~\ref{fig:RM_TS_Ha} shows that the magnitude of the RM effect on the \Ha\ line is about 0.3$\%$ in the blue part of the line (around 30$\%$ of the line core contrast), and about $\lesssim0.05$--$0.1\%$ in the other parts of the line. The RM effect correction is made down to the noise level. We note that our results for \Ha\ are in conflict with previous literature: by $\sim$3~$\sigma$ with \citet[who used the CARMENES spectrograph][]{Yan2018}, by $\sim$4.5~$\sigma$ with \citet[][PEPSI]{Cauley2018} and by $\sim$1~$\sigma$ with \citet[][CARMENES]{Turner2020}. The difference could come from our new extraction method, the different resolutions, or different normalizations. However, we note that \citet{Casasayas-Barris2019} reported \Ha\ detections in KELT-20~b/MASCARA-2~b, that are different for the HARPS-N and CARMENES instruments, despite using the same data reduction. It is thus likely that there are instrumental systematics that are not yet taken into account (detector non-linearity, blaze and bias removal, spectrum interpolations, moonlight pollution, etc.). A follow-up study exploring such effects must be undertaken.

\item \Hb: The \Hb\ line and its trail are also clearly detected (16.1~$\sigma$, Fig.\ref{fig:Ha_Hb_lines}). Our \Hb\ detection also disagrees with the result of \citet[][by $\sim$2~$\sigma$]{Cauley2018} and is subject to the same caution. Another explanation for the \Ha\ and \Hb\ differences between studies could be intrinsic variation in the atmosphere. We also note that the line absorption is stronger in the second part of the transit than in the first part, hence the average measured line could arise from different weights given to each spectrum (due to the observation conditions).

\item \Hg: For \Hg\, the atmospheric trail is barely visible, since the total spectrum 7.5~$\sigma$ detection is distributed to $\sim$1~$\sigma$ in each spectrum (Fig.\ref{fig:Hg_Hd_lines}). \Hb\ and \Hg\ have only been reported in the atmosphere of three exoplanets: HD189733~b, KELT-9~b, and KELT-20~b/MASCARA-2~b \citep{Cauley2016,Cauley2018,Casasayas-Barris2019}.

\item \Hd: The \Hd\ detection in the KELT-9~b atmosphere (4.6~$\sigma$) is the first of its kind to our knowledge (Fig.\ref{fig:Hg_Hd_lines}). Despite its significance, the line is probably slightly biased because the local stellar signal becomes less visible and is more difficult to correct for. This may be due to the lack of signal in the bluer part of the detector.

\item \He: The \He\ absorption (2.9~$\sigma$) is just below the significance level of 3~$\sigma$. Therefore it should be treated as a hint of detection (Fig.\ref{fig:He_Hz_lines}). Additional data are needed to better constrain the \Hd\ and \He\ line shapes.

\item \Hz: \Hz\ is not detected in our data set (Fig.\ref{fig:He_Hz_lines}). The most probable reason is due to the lack of flux in the very blue part of the HARPS-N wavelength range. Our upper limit value is hinting that the \Hz\ absorption is smaller than all the other Balmer lines, as expected.
\end{itemize}

Table~\ref{tab:line_fit} presents a summary of our spectral analysis of the Balmer series. From the contrasts of the KELT-9~b Balmer \Ha, \Hb, and \Hg\ lines, we know that neutral hydrogen is present at altitudes between $\sim1.2~{\rm R_P}$ and $\sim1.5~{\rm R_P}$ (see Fig~\ref{fig:Ha_Hb_lines} and~\ref{fig:Hg_Hd_lines}). Since this is close to the planetary Roche radius ($1.95\pm0.2$~${\rm R_P}$), the strong Balmer lines absorptions are a hint of atmospheric escape \citep{Yan2018}. We note that no line is significantly blueshifted or redshifted in the line of sight. With an average shift of 0.2$\pm$0.6~\kms, we do not infer any strong winds in the upper atmosphere. This is compatible with the high circulation drag expected in very hot atmospheres \citep[][but also see \citet{Wong2019} who found a relatively high recirculation efficiency]{Koll2018}. Finally, we note that the \Ha, \Hb, and \Hg\ lines have a similar FWHM, with an averaged value of $44.9\pm1.4$~\kms. For the rest of our analysis, we investigate the strong and consistent \Ha, \Hb, and \Hg\ signals that contain most of the information.

\begin{table}
\caption{Summary of the hydrogen Balmer line detections.}
\begin{center}
\begin{tabular}{lcccc}
\hline
\rule[0mm]{0mm}{5mm} & C & $v$ & FWHM & $\delta$\tablefootmark{\textasteriskcentered}\\
& [\%] & [\kms] & [\kms] & [\%]\\
\hline
\Ha & $0.91\pm0.04$ & $2.1\pm0.8$ & $44.3\pm1.8$ & $0.68\pm0.03$\\
\Hb & $0.68\pm0.04$ & $-1.7\pm1.3$ & $46.3\pm2.5$ & $0.45\pm0.03$\\
\Hg & $0.44\pm0.07$ & $0.9\pm2.9$ & $43.5\pm5.5$ & $0.31\pm0.04$\\
\Hd & $0.61\pm0.13$ & $-4.6\pm2.2$ & $25.6\pm5.7$ & $0.42\pm0.09$\\
\He & $0.39\pm0.14$ & $-15.0\pm5.0$ & $29.5\pm10.0$ & $0.26\pm0.09$\tablefootmark{\dag}\\
\Hz & $\leq0.50$ & - & - & $\leq0.34$\tablefootmark{\dag}\\
\hline
\end{tabular}
\end{center}
\tablefoot{}
\tablefoottext{\textasteriskcentered}{The absorption depth is averaged over the line full width \citep[see][]{Wyttenbach2017}, takes systematic noise into account, and is used to compute the line detection significance.}
\tablefoottext{\dag}{\He\ is not statistically detected. For \Hz\, we show the $3\sigma$ upper limit assuming a Gaussian line shape with a fixed $v=0$~\kms\ and ${\rm FWHM}=30$~\kms.}
\label{tab:line_fit}
\end{table}

%%%%%%%%%%%%%%%%%%%%%%%%%%%%%%%%%%%%%%%%%%%%%%%%%%%%%%%%%%%%%%%
\section{Interpretation of the Balmer lines of KELT-9~b}\label{Sec_inter}
%%%%%%%%%%%%%%%%%%%%%%%%%%%%%%%%%%%%%%%%%%%%%%%%%%%%%%%%%%%%%%%

The measured FWHMs of the KELT-9~b Balmer \Ha, \Hb, and \Hg\ lines cannot be explained only by thermal broadening. Indeed, for hydrogen, such a FWHM would require a thermal broadening with a temperature around 40\,000~K, far above any theoretical prediction for a thermosphere temperature \citep[e.g.,][]{Salz2016}. Since the temperature cannot be measured directly, a rough estimate of the atmospheric properties of KELT-9~b from the Balmer series can be made following \citet[][section 2]{Huang2017} as done in previous studies \citep{Yan2018,Turner2020}. The estimation of \citet{Huang2017} assumes an isothermal hydrostatic atmosphere with a mean molecular weight $\mu$=1.3 and a temperature $T$=5000~K. For KELT-9~b, a better assumption is an atmosphere dominated by ionized hydrogen ($\mu$=0.66) and a thermosphere temperature of $T$=10\,000~K \citep{Fossati2018,GarciaMunoz2019}. With these numbers, we find that to explain the FWHM\footnote{To compare our FWHM, we removed the additional width contribution by other effects (see Sec.~\ref{sec:pawn_model} and~\ref{sec:LTE}).} of our \Ha\ detection, one needs an optically thick layer ($\tau$$\sim$40) of excited hydrogen \ion{H}{i}(2) with a number density of n$\sim$$8.2\times10^3$~cm$^{-3}$ along a typical line path of L$=$$2\sqrt{2{\rm R_P N_{sc} H}}$$\sim$13.8$\times10^9$~cm, where ${\rm N_{sc}}$$\sim$6 is the number of absorbing scale-heights \citep{Madhusudhan2014c,Madhusudhan2015,Huang2017}. This means that at an altitude of ${\rm N_{sc} H}\sim0.3~{\rm R_P}$ ($H\sim6\,000$~km), which is about half of the line core contrast, the \Ha\ line core have $\tau\gg1$. Since the line is optically thick in most of the layers around that altitude, the line FWHM is larger than the thermal width \citep{Huang2017}. Also, this is consistent with our observations since the \Ha\ core is observed at $\sim1.5{\rm R_P}$, where $\tau$ becomes close to unity \citep{Lecavelier2008,DeWit2013,Heng2015,Heng2017b}.

Another estimate of the atmospheric conditions can be done with the \citet{Lecavelier2008} formalism, which again assumes an isothermal hydrostatic atmosphere. Indeed, one can measure the atmospheric scale-height $H$ thanks to the linear correlation between the line contrasts (or the altitudes $z$) and the hydrogen Balmer line oscillator strengths (see the $\log_{10} gf$ values in Sec.~\ref{sec:pawn_model}). Knowing that $\Delta z = H\Delta \ln(gf) = H\ln(10)\Delta\log_{10}(gf)$, the slope of the $z$--versus--$\ln(gf)$ linear fit gives us $H$. We computed that $H=9\,600\pm1\,400$~km around the average altitude of absorption (1.3 ${\rm R_P}$). When the gravity is evaluated at 1.3 ${\rm R_P}$ ($\sim0.59\,g_{\rm surf}$), we have that $T/\mu\simeq13\,500\pm2\,000$~[K/u]. Assuming a $\mu$ between 0.66 and 1.26 (atmospheric composition dominated by a mixture of ionized and neutral hydrogen and helium), the upper atmosphere temperature is constrained between 7\,500 and 19\,500 K. The lower half temperature range is more probable because the thermal ionization of hydrogen decreases $\mu$\footnote{A composition dominated by molecular hydrogen and helium ($\mu=2.34$) is probably excluded by an unphysical T$\gtrsim$ 30\,000 K.}.

Finally, our line contrast measurements and the aforementioned assumptions lead to the same conclusions as in \citet{Yan2018} of a Jeans mass loss rate of $\sim$10$^{12}$~g\,s$^{-1}$, although hydrodynamic escape is more likely the mass loss mechanism in KELT-9~b \citep{Fossati2018}.

While this kind of estimation is useful for getting a qualitative idea of the planet aeronomic conditions, the data quality calls for a quantitative estimate. A few forward self-consistent models of the hydrogen \Ha\ line exist \citep{Christie2013,Huang2017,Allan2019,GarciaMunoz2019}. They explore different physical questions and have different atmospheric structures, but none of them are linked to any retrieval algorithms. Retrieval approaches (on similar absorption lines) have been made by \citet{Cauley2018,Fisher2019,Seidel2020}, but without the mass loss rate being a parameter. Thus, we decided to follow our own retrieval approach with a model adapted to our data set.

%--------------------------------------------------------------------------------------------------------------------------------------------------------------------------
\subsection{The \texttt{PAWN} model}\label{sec:pawn_model}
%--------------------------------------------------------------------------------------------------------------------------------------------------------------------------

We model the planet atmosphere in transmission with the customized model \texttt{PAWN} (PArker Winds and Saha-BoltzmanN atmospheric model) which is included in our \texttt{CHESS} framework as a tool for modeling exoplanetary thermospheric lines in transmission spectroscopy. Our purpose is to constrain fundamental parameters of the thermosphere regions such as the temperature, the mass loss rate and the hydrogen number density (local thermodynamical condition), if possible. Another goal is to have a sufficiently fast model to feed a retrieval algorithm. Hence, our modeling holds several assumptions.

We model a 1-D planetary atmosphere, with a special interest in the upper atmosphere part (thermosphere). A logarithmic grid with $n_r$ radius $r$ layers is chosen between 1 and ${\rm r_{up}~R_P}$. For the hydrogen Balmer lines, ${\rm r_{up}}$=2 ensures the convergence of the models. We model the atmospheric structure $\rho(r)$ with a hydrostatic or a hydrodynamic structure. For the hydrostatic solution, we have the base density $\rho_0(r\!=\!1)$ as a free parameter, and consider that the pressure scale-height $H=k_BT/\mu g$ can change with the altitude according to $T(r),\ \mu(r),\ g(r)$ \citep[see e.g., the $^\pi\eta$ code in][]{Ehrenreich2006,Pino2018a,Seidel2020}. In the hydrostatic case, we can estimate a lower limit to the mass loss rate, by computing the Jeans escape rate at the planetary Roche radius, ${\rm r_{Roche}}$ \citep{Eggleton1983,RiddenHarper2016}, or at the exobase radius, ${\rm r_{exobase}}$, if below \citep{Lecavelier2004,Tian2005,Salz2016,Heng2017}:

\begin{equation} \label{eq:MLR_J}
\dot{{\rm M}}_{\rm Jeans} = 4\pi {\rm r^2_{Roche}} \mu(r_{Roche}) F_{\rm Jeans}(r_{Roche}),
\end{equation}

where $F_{\rm Jeans}$ is the Jeans escape flux \citep{Shizgal1996,Hunten1989,Chamberlain1987}. For the hydrodynamical solution, we use the isothermal Parker wind solution following the formalism of \citet{Heng2017}\footnote{
Eq. 13.14 of \citet[][Chapter 13]{Heng2017} should be corrected as follows:
\begin{equation*} \label{eq:MLR_J}
\mathcal{M}^2-{\rm ln}(\mathcal{M}^2)-1=4\,{\rm ln}(x)+4\,(1/x-1),
\end{equation*}
where the Mach number $\mathcal{M}=v/c_s$, and $x=r/r_s$, with $c_s$, $r_s$ being the sound speed, and the sonic point radius, respectively.
}, \citet{Oklopcic2018}, and \citet{Lampon2020}. Our free parameter for the atmospheric structure becomes the atmospheric mass loss rate: 

\begin{equation} \label{eq:MLR_PW}
\dot{{\rm M}}_{\rm PW} = 4\pi r^2 \rho(r) v(r),
\end{equation}

where $\rho(r)$ and $v(r)$ solve the Parker equations \citep{Parker1958,Lamers1999} from the sonic point ($r_s,\ v_s,\ \rho_s,\ \mu_s$) down to the lowest radius in the grid. We note that $\dot{{\rm M}}_{\rm PW}$ is constant through the atmosphere. Since the Parker wind solution is isothermal, 
we keep an isothermal profile for both the hydrostatic and hydrodynamic structure with the temperature $T$ as a free parameter. We note that the assumption of an isothermal profile for the upper thermosphere (typically between 1.1 and 2--3 ${\rm R_P}$ for hot Jupiters) is a good approximation of aeronomy models \citep[e.g.,][]{Lammer2003,Lecavelier2004,Yelle2004,GarciaMunoz2007,MurrayClay2009,Koskinen2013a,Salz2016}. Indeed, above the thermobase where the temperature changes dramatically, the temperature encounters a maximum and is varying much slower through several orders of magnitude in pressure than in the lower regions.

For the atmospheric constituent structure, we use a pre-computed chemical equilibrium table obtained with the chemical equilibrium code of \citet{Molliere2017}. The chemical equilibrium (of a variety of neutral and singly ionized atomic, molecular, and condensed species) is computed, with a Gibbs energy minimization, into a grid of pressure ($10^{-15}$--$10^2$~bar) and temperature (1000--20000~K). The mean molecular weight $\mu(\rho(r),T(r))$, the mass fraction $m_\iota(r)$, and volume mixing ratio $n_\iota(r)$ are given by the chemical grid, where $\iota$ is a chosen element and ionization of interest (\ion{H}{i}, \ion{He}{i}, \ion{Na}{i}, \ion{K}{i}, \ion{Ca}{ii}, etc.). For example, in the hydrogen case, we know the ratio of ionized to neutral hydrogen $n_\ion{H}{ii}/n_\ion{H}{i}$ from reading and interpolating the chemical grid with the correct $T$ and $\rho$ from our atmospheric structure. When the Parker wind solution is in use, we link it with the chemical grid at the sonic point down to the lowest radius. We note that for hydrogen, the Gibbs energy minimization is equivalent to computing the Saha law with an electron number density $n_e$ coming from the ionization of the hydrogen and from all the other atomic species in the grid. By default, the chemical grid is always used, but it is also possible to assume a constant (free parameter) $\mu$ through the atmosphere. In that case the $n_\ion{H}{ii}/n_\ion{H}{i}$ can be constant or can follow the Saha equation with $n_e$ accounting only for the electron coming from the hydrogen ionization. In any case, we are ignoring photo-ionization effects that can be important in the upper atmosphere.

Once we know the vertical distribution of our element of interest, we still need to know what is the number density of the electronic level $\iota(n)$ from which the transitions we observed are arising. In the case of \ion{H}{i}, we are observing the Balmer series in absorption, which arise from photons being absorbed by the n=2 electronic state, that is to say the \ion{H}{i}(2) state. If the atmosphere is in local thermodynamic equilibrium (LTE), the distribution of the electronic states follow the Boltzmann equation (for hydrogen):

\begin{equation} \label{eq:MLR}
\frac{n_{\ion{H}{i}(n)}}{n_{\ion{H}{i}}}=\frac{2 n^2}{\upsilon(T)}e^{E_1/k_BT\cdot(n^2-1)/n^2},
\end{equation}

where $E_1$=13.606 eV, and $\upsilon(T)$ is the hydrogen partition function. The Boltzmann equilibrium is assumed by default in our code. However, because we are probing high in the thermosphere, the collisions between particles become less numerous and the electronic state distribution can depart from the Boltzmann distribution, the atmosphere is in non-LTE or NLTE \citep{Barman2002,Fortney2003,Christie2013,Huang2017,Oklopcic2018,Fisher2019,Allan2019,GarciaMunoz2019,Lampon2020}. For the NLTE case, we define the departure coefficients \citep[e.g.,][]{Mihalas1970,Salem1979,Barman2002,Draine2011,Hubeny2014} as:

\begin{equation} \label{eq:MLR}
\beta_{\ion{H}{i}(n)}=\frac{n_{\ion{H}{i}(n)}^{NLTE}}{n_{\ion{H}{i}(n)}^{LTE}}.
\end{equation}

Then, we can let the different $\beta_{\ion{H}{i}(n)}$ be free parameters (or equivalently all the $n_{\ion{H}{i}(n)}$ are free parameters). Another possibility is to force all the $n_{\ion{H}{i}(n)}$ to depart by the same amount from LTE, meaning that only one constant free parameter $\beta$ is used for all $n_{\ion{H}{i}(n)}$. We note that a constant $n_{\ion{H}{i}(n)}/n_{\ion{H}{i}}$ through the thermosphere is an assumption justified by \citet{Huang2017} and \citet{GarciaMunoz2019}. 

When an element and its transitions are chosen (in our case the Hydrogen Balmer series), the opacity $\kappa$~[cm$^2$g$^{-1}$] are computed in each layer of the atmosphere according to the line list of \citet{Kurucz1979,Kurucz1992}, to the temperature, and to the number density $n_{\ion{H}{i}(n)}$. We do assume that the line intensity arises from photon absorption and from stimulated emission \citep{Sharp2007}. For the hydrogen Balmer transition from electronic level i to j, we have:

\begin{equation} \label{eq:opa}
\kappa_{ij} = \frac{\pi e^2}{m_e c} \frac{g_i f_{ij}}{m_{\rm H} n_{\ion{H}{i}}} \left(\frac{n_{\ion{H}{i}(i)}}{g_i} - \frac{n_{\ion{H}{i}(j)}}{g_j}\right) \Phi_{\rm V}(\nu_{ij}),
\end{equation}

where $e$ is the elementary charge, $m_e$ is the electron mass, and $c$ the speed of light in vacuum. The $f_{ij}$, $g_i$, $g_j$ are the oscillator strength, and the statistical weights of the electronic levels due to their degeneracies, respectively. For the hydrogen Balmer series, we have $\log_{10} g_2 f_{2j}$=0.71, -0.02, and -0.447 for $j$=3, 4, and 5, respectively. A Voigt profile $\Phi_{\rm V}$, centered on the line transition center $\nu_{ij}$ (or $\lambda_{ij}$), is assumed for the line shape. The Lorentzian wing component considers the natural broadening (for the hydrogen Balmer series, the Einstein A-coefficients are $\log_{10}(A_{j2}~[s^{-1}])$=8.76, 8.78, 8.79 for $j$=3, 4, and 5, respectively). There is no pressure broadening. The Gaussian core component takes the thermal broadening into account. We can also add velocity components due to rotation, circulation, winds in the atmosphere (see below). For this exploratory study, we decided not to consider the opacity arising from H$^-$ \citep[the main source of background continuum opacity,][]{Kitzmann2018}, as a rough estimate shows that its opacity ($\sim$10$^{2}$~cm$^2$g$^{-1}$) stays smaller than the one from each Balmer line ($10^3$--$10^8$~cm$^2$g$^{-1}$ in the Gaussian core). Hence, the modeled absorptions are generally robust for the line cores that we observe \citep{Hoeijmakers2019}. The spectral resolution of our model is chosen to be 0.001~$\AA$ ($\lambda/\Delta\lambda\sim10^6$), and the wavelength coverage is 20~$\AA$ around each line.

Next, we perform the radiative transfer in a grazing geometry following the method of \citet{Ehrenreich2006,Molliere2019}. For each impact parameter, the optical depth $\tau=\sum\kappa\tilde{m_\iota}$, where $\tilde{m_\iota}$ is the column mass of the element $\iota$ in [g~cm$^{-2}$], is computed through the line of sight (LOS). The transmission function $\mathcal{T}=e^{-\tau}$ is computed according to a pure absorption in each LOS. The transmission spectrum is computed assuming a spherically symmetric atmosphere, weighted along the impact parameter surface annuli, and normalized to the stellar surface.

Within the previous two steps, our line profile can be broadened according to different atmospheric patterns \citep[see][]{Brogi2016,Seidel2020}. First, we assume a solid body planetary rotation (perpendicular to the orbital plane). Since the planet rotation is axisymmetric, this breaks the spherical symmetry assumed before. In that case, we divide each hemisphere of the planet in a grid of $m_\theta$ circular sector. The sector grid is evenly separated in $\cos\theta$, where $\theta$ is the angle going from the equator to the pole back to the equator (anti-clockwise). If the rotation is included, our code becomes essentially a 2-D atmospheric code. The planet rotation period ${\rm P_P}$ is the parameter that defines the solid rotation. ${\rm P_P}$ can be a free parameter, or more simply be fixed to the planetary orbital period by assuming that the planet is tidally locked (which is our default choice). It is interesting to note that in this configuration, each LOS has a fixed velocity along the cord, which is given by:

\begin{equation} \label{eq:MLR}
v^{\rm LOS}_{\rm rot}(r,\theta)=\frac{2\pi r}{\rm P_P} \cos(\theta).
\end{equation}

In the case of KELT-9~b, the planetary rotation have a broadening effect of $\sim6.1$~\kms\ at 1 ${\rm R_P}$ for each hemisphere ($\sim12.2$~\kms\ in total).

Second, when the atmosphere is hydrodynamical (Parker winds), a spherically symmetric expansion of the atmosphere occurs with a velocity $v(r)$. This expansion changes the projected velocity of the different components of each LOS, thus creating another broadening source. This Parker wind broadening can be considered in our model, but not simultaneously to the rotation (because the symmetry is not the same, the structure would be 3-D and the computational cost would become too large). The LOS velocity is given by:

\begin{equation} \label{eq:MLR}
v^{\rm LOS}_{\rm PW}(r_i,r_j)=\frac{v(r_j)}{r_j}\sqrt{r_j^2-r_i^2}.
\end{equation}

In practice, for KELT-9b, for most of the mass loss rates (except for unphysical large rates), the upward velocity $v(r)$ is smaller than 0.01--1~\kms\ in the probed layers (i.e., the sonic point $r_s$, where the velocity becomes several \kms\, is far above the Roche radius). This leads to a negligible (much smaller than the instrumental resolution) parker wind broadening \citep[though see][for other possible scenarios]{Seidel2020}. Thus, for the rest of our study, we decided to ignore this effect in favor of the planetary rotation. We note that these two aspects have been previously explored in \citet{Seidel2020}. They were able to keep a 1D atmospheric structure at the price of several approximations in the atmospheric velocity patterns (pseudo-2D/3D).

In the next step, our line profile is modified according to the observational set-up. First, the orbital blurring due to the planet changing velocity during an exposure time is computed with a box-shape convolution (of a width of $\sim$6.8~\kms\ for the KELT-9~b orbit and for the 600~s exposure time). The spectrum is then convolved with the instrument profile, which is assumed to a be a Gaussian (with a width of $\sim$2.7~\kms\ for HARPS-N). Then, the spectrum is binned down to match the data sampling. Finally, the differential normalization \citep[due to the loss of the absolute depth in ground-based observations, see][]{Wyttenbach2015,Wyttenbach2017,Pino2018a} is performed according to the aforementioned reference passbands of each line. The models, as for the data, are normalized to the white light curve transit depth of KELT-9b ($\delta$=0.00677). The models are now in a form which is comparable with the data. Our \texttt{PAWN} model is finally linked with the Markov Chain Monte Carlo (MCMC) posterior sampling code \texttt{emcee} \citep{Foreman-Mackey2013}. Typically, for one line, one model without rotation is computed in $\lesssim$0.1~s ($n_r$=100), while with rotation, one model takes $\sim$0.5~s ($n_r$=50, $m_\theta$=14). We note that a model with $n_r$=50 and $m_\theta$=14 is $\lesssim$1\% equal to a model with $n_r$=250 and $m_\theta$=28.

In the next section, we apply our \texttt{PAWN} model to the Balmer lines detection in the KELT-9~b atmosphere.

%--------------------------------------------------------------------------------------------------------------------------------------------------------------------------
\subsection{Interpretation of the Balmer series in the KELT-9~b atmosphere with the \texttt{PAWN} model}\label{sec:pawn_application}
%--------------------------------------------------------------------------------------------------------------------------------------------------------------------------

In this section, we investigate the \Ha, \Hb, and \Hg\ signals based on their line shape. All lines are fitted together in a range of 20~$\AA$ (6000 data points, notably to cover the normalization passbands). Our retrieval approach uses our \texttt{PAWN} model with the \texttt{emcee} sampler to explore the parameter space and to find the best models and their parameter confidence intervals (CI). For each different set of models, we use ten walkers for each dimension (parameter) during 2500 steps and with a burn-in size of 500 steps. We use the bayesian information criterion (BIC) to compare between models.

For every model, each line center position is a free parameter (three parameters for \Ha, \Hb, and \Hg). The line center positions are sensitive to circulation winds in the upper atmosphere \citep{Snellen2010,Brogi2016,Seidel2020}. Uniform priors between $\pm$20~\kms\ are used to ensure that our prior cover the full posteriors. We noticed a posteriori that these parameters are not correlated with any of the other parameters. Also, we found that our MCMC results are confirming the line shifts and errors quoted in Table~\ref{tab:line_fit}, thus we do not show the results for these parameters. The second set of parameters defines the atmospheric structure with the temperature $T$, and the base density $\rho_0(r\!=\!1)$ or the mass loss rate $\dot{{\rm M}}$ depending on if the structure is hydrostatic or hydrodynamic, respectively (for these latter, we use their $\log_{10}$ values). We use a uniform prior on T between 1000 and 20000~K. We use a log-uniform prior between -18 and -4 for the $\log_{10}(\rho_0\,[{\rm g\,cm}^{-3}])$. We use a log-uniform prior between 8 and 17 for the $\log_{10}(\dot{{\rm M}}\,[{\rm g\,s}^{-1}])$. In our base models, we use the Boltzmann equilibrium and there are no other parameters. We first explore this set-up, compare the hydrostatic and hydrodynamic models, before exploring some NLTE models.

%--------------------------------------------------------------------------------------------------------------------------------------------------------------------------
\subsubsection{LTE models with a hydrostatic or hydrodynamic structure}\label{sec:LTE}
%--------------------------------------------------------------------------------------------------------------------------------------------------------------------------

\begin{figure}[t!]
\centering
\includegraphics[width=0.47\textwidth]{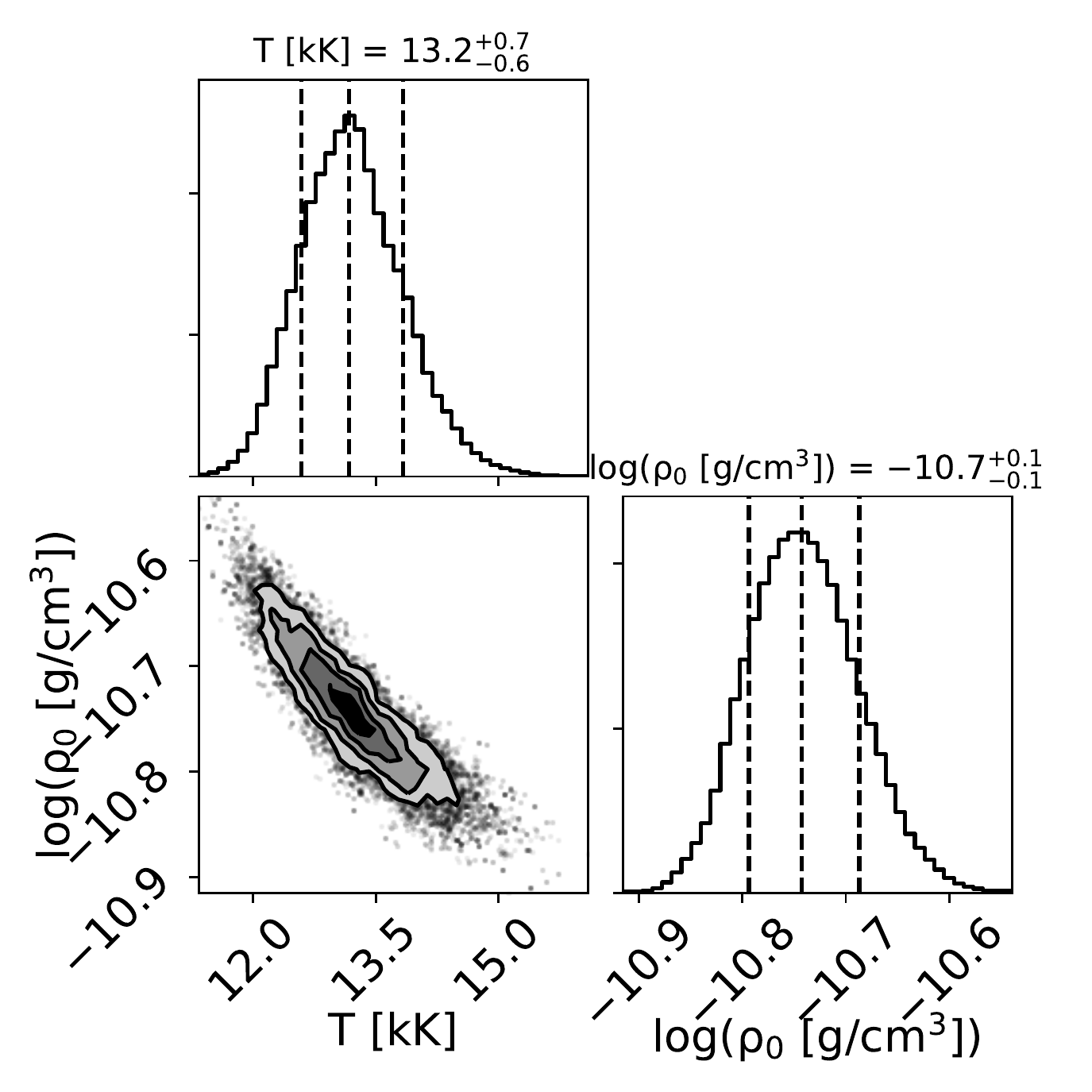}
\caption{Correlation diagrams in the case of a hydrostatic atmosphere in the Boltzmann equilibrium.}
\label{fig:pawn_HS_SB}
\end{figure}

\begin{figure}[t!]
\centering
\includegraphics[width=0.47\textwidth]{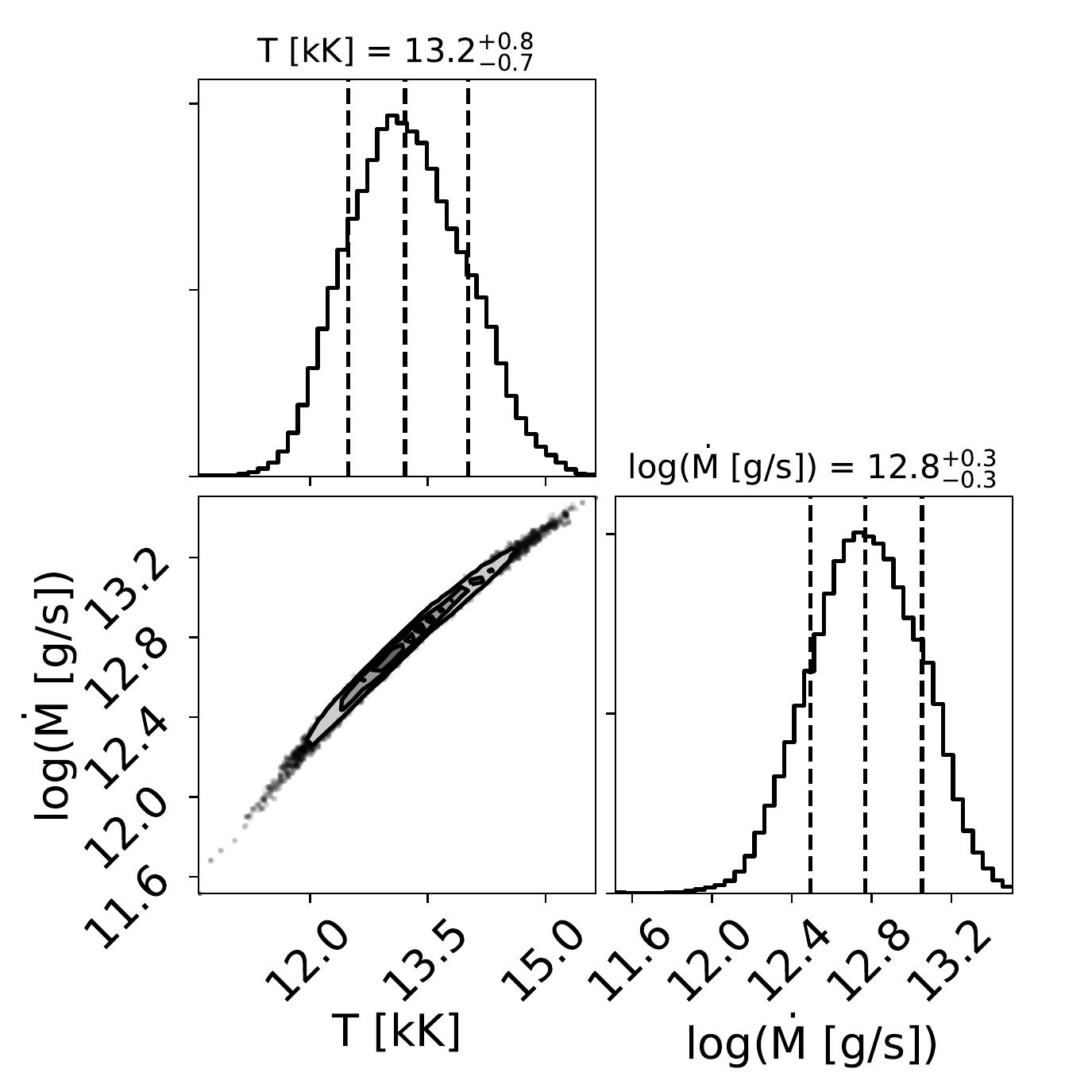}
\caption{Correlation diagrams in the case of a hydrodynamic atmosphere in the Boltzmann equilibrium.}
\label{fig:pawn_PW_SB}
\end{figure}

We first asses how our simplest models are able to fit the observed lines. These two set of models are hydrostatic and hydrodynamic, respectively. They have two free parameters (apart of the line centers), and assume a Boltzmann equilibrium. For the hydrostatic model, we find a solution with a thermosphere temperature of $T=13\,180^{+650}_{-580}$~K, and a base density of $\log_{10}(\rho_0\,[{\rm g\,cm}^{-3}])=-10.74\pm0.05$ (BIC=5016.1, Fig.~\ref{fig:pawn_HS_SB}). The corresponding Jeans mass loss rate at the Roche radius is computed to be ${\rm \dot{M}}_{\rm Jeans}\sim6.6\times10^{11}$~g\,s$^{-1}$. Interestingly, our hydrodynamical model, which has a BIC=5017.2, is found to have a similar thermosphere temperature of $T=13\,200^{+800}_{-720}$~K, while the mass loss rate is tending towards to the higher value of $\log_{10}({\rm \dot{M}}_{\rm PW}\,[{\rm g\,s}^{-1}])=12.8\pm0.3$ (i.e. ${\rm \dot{M}}_{\rm PW}\simeq5.9\times10^{12}$~g\,s$^{-1}$, Fig.~\ref{fig:pawn_PW_SB}). Both models are comparable in term of statistics and none can be preferred; this is because the density profiles are not different enough at the altitudes probed. The mass loss rate is in both cases around 10$^{12}$~g\,s$^{-1}$, confirming the result of \citet{Yan2018}. Also, we can safely conclude that the thermosphere temperature, if in LTE, must be near 13\,000~K (in line with our first estimate using the \citet{Lecavelier2008} equation). Despite the high temperature, the line FWHMs ($44.9\pm1.4$~\kms) are not dominated by thermal broadening. Indeed, for $T=13\,200$~K, the hydrogen thermal broadening is $\sim25$~\kms, while the other source of broadening are $\sim16$~\kms, $\sim7$~\kms, and $\sim2.7$~\kms\ for the rotational (at $r=1.3{\rm R_P}$), orbital blurring, and instrumental broadening, respectively. This means that the first source of broadening ($\sim33$~\kms) comes from optically thick absorptions through most of the lower layers \citep{Huang2017}.

We notice that the line contrasts and FWHMs are both enhanced when the temperature or the mass loss rate are increasing. For the temperature, this is caused by the atmosphere being more extended, and by thermal broadening. For the mass loss rate, this is caused by larger densities present at higher altitudes that increase the LOS optical depth, thus boosting the broadening and the contrast. Therefore, the shape of the correlation in the hydrodynamical case (Fig.~\ref{fig:pawn_PW_SB}) is explained by the following mechanism:\ when the temperature increases, the number density of the hydrogen second electronic level builds up. However, as for the stellar case, the total number density of neutral hydrogen decreases because of ionization. This last effect eventually dominates at high temperature. So, to keep a constant absorption corresponding to our observations, a higher density is needed. This is achieved by increasing the mass loss rate (Eq.~\ref{eq:MLR_PW}). Thus, we get the positive correlation between $T$ and ${\rm \dot{M}}_{\rm PW}$, which shape is given by the Boltzmann equation.

%--------------------------------------------------------------------------------------------------------------------------------------------------------------------------
\subsubsection{NLTE models with a hydrodynamic structure}\label{sec:NLTE}
%--------------------------------------------------------------------------------------------------------------------------------------------------------------------------

We explore the NLTE case for the hydrodynamical atmosphere case. We first run MCMC chains by letting all the electronic level number density be free parameters. The prior on each $\log_{10}(\beta_{\ion{H}{i}(n)}n^{LTE}_{\ion{H}{i}(n)}/n_{\ion{H}{i}})$ was set to be a log-uniform distribution between -40 and 0. With this test, we found that the model hardly converges, and that the parameters are not well-constrained (see Fig.~\ref{fig:pawn_PW_noSB_free_all}). The explanation lies in two different families of solutions. One family is distributed around the Boltzmann equilibrium, while the other is away from it, and has a lower temperature and a higher mass loss rate. We therefore further explore these two families.

\begin{figure}[t!]
\centering
\includegraphics[width=0.47\textwidth]{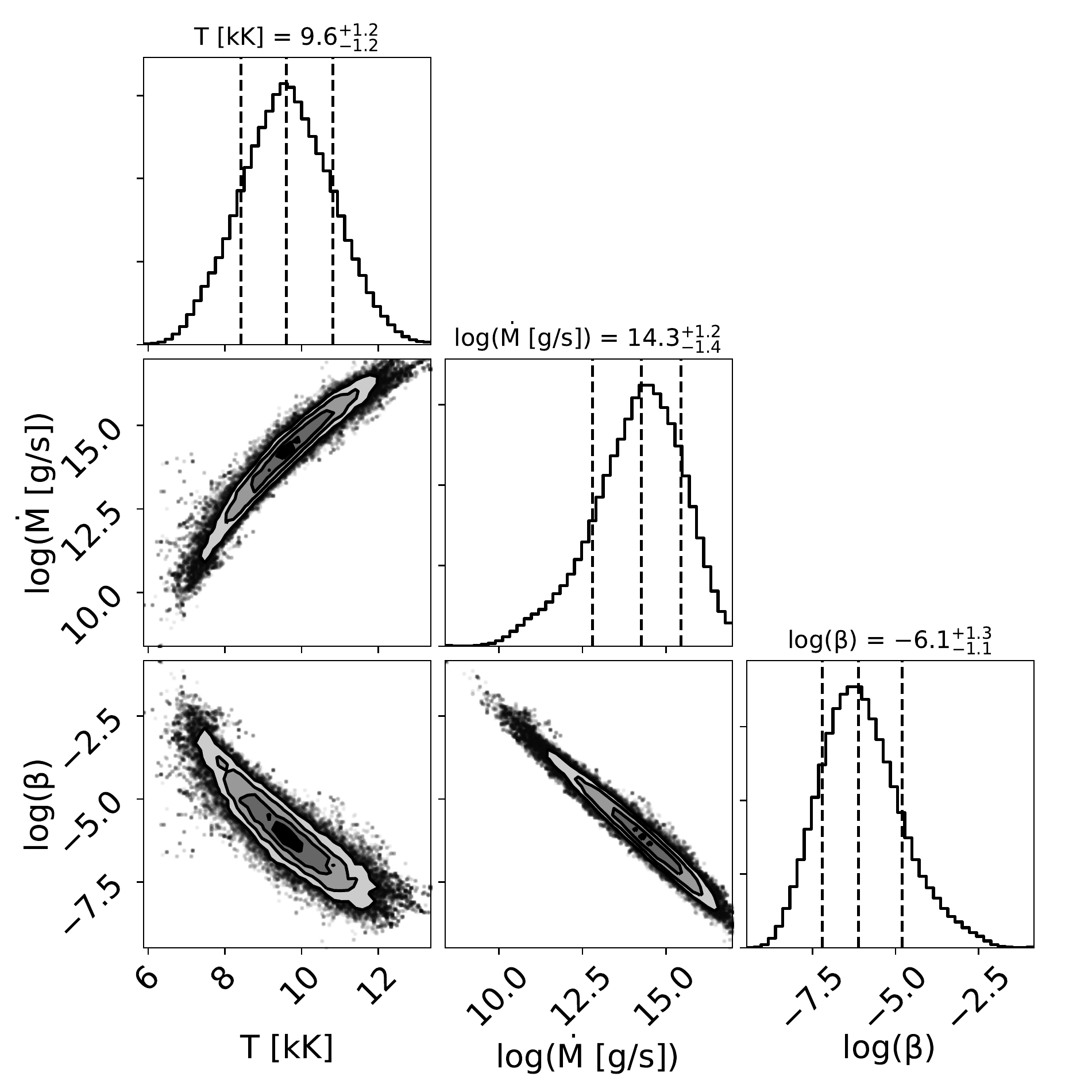}
\caption{Correlation diagrams in the case of a hydrodynamic atmosphere with a departure $\beta$ from LTE.}
\label{fig:pawn_PW_noSB_dep}
\end{figure}

For the first family, we limit our prior ranges on the $\log_{10}(\beta_{\ion{H}{i}(n)}n^{LTE}_{\ion{H}{i}(n)}/n_{\ion{H}{i}})$ between -9, -12,-16, -20 and 0, for n=2, 3, 4, 5, respectively. The results are shown in Fig.~\ref{fig:pawn_PW_noSB_free}, and the model results in a BIC=5031.4. The increased number of parameters clearly exclude this model ($\Delta{\rm BIC}$=-14.2<-5), despite a better best fit. We still notice that the retrieved temperature lies between 12\,600 and 16\,600~K, while the mass loss rate lies between $10^{12}$ and $10^{16}$~g\,s$^{-1}$, encompassing the LTE result.

\begin{figure*}[t!]
\centering
\includegraphics[width=0.97\textwidth]{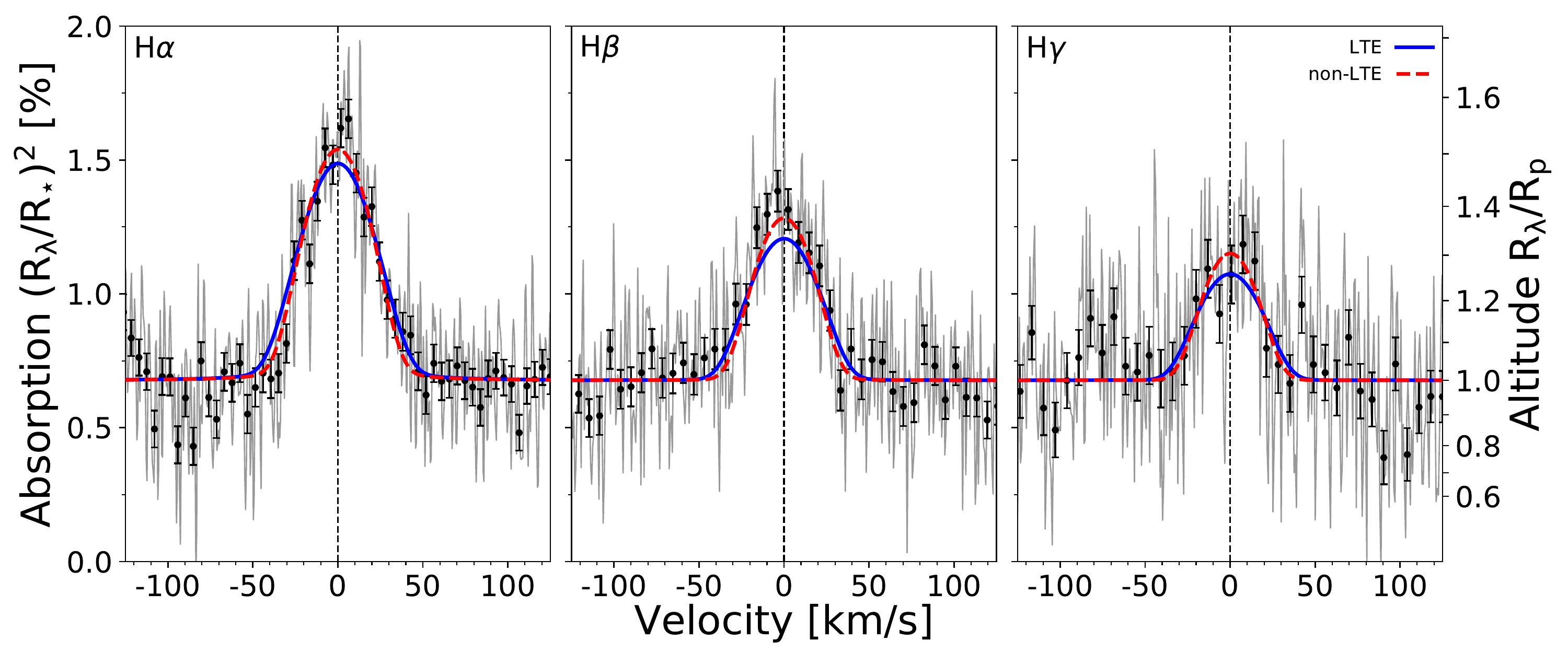}
\caption{\Ha, \Hb, and \Hg\ transmission spectra (light gray, and binned by $10\times$ in black circles) in the planet rest frame and in velocity space (the continuums are set to the white light transit depth $\delta=0.00677$). The best \texttt{PAWN} models, in the case of a hydrodynamically expanding thermosphere, and in the Boltzmann equilibrium, are shown in blue. The retrieved temperature is $T=13\,200^{+800}_{-720}$~K, and the mass loss rate is $\dot{{\rm M}}_{\rm PW}=10^{12.8\pm0.3}$~g\,s$^{-1}$. The best \texttt{PAWN} models, in the case of a hydrodynamically expanding thermosphere, with a LTE departure, are shown in dashed red. The retrieved temperature is $T=9\,600^{+1210}_{-1180}$~K, the mass loss rate is $\dot{{\rm M}}_{\rm PW}=10^{14.3^{+1.2}_{-1.4}}$~g\,s$^{-1}$, and the common LTE departure coefficient is $\beta=10^{-6.1^{+1.3}_{-1.1}}$.}
\label{fig:best_fit_PW_SB}
\end{figure*}

For the second family of solutions in NLTE, we notice that it is mainly dominated by the departure of the second electronic level. We explore this family, by reducing the number of parameters for the number densities. Instead of having all the $\log_{10}(\beta_{\ion{H}{i}(n)}n^{LTE}_{\ion{H}{i}(n)}/n_{\ion{H}{i}})$ be free parameters, we decide to use a unique departure coefficient $\beta$ for all the electronic levels, that is $\beta_{\ion{H}{i}(n)}=\beta$ for all {\ion{H}{i}(n)}. For this last MCMC run, a log-uniform prior for $\log_{10}(\beta)$ was set between -10 and 5. The result is shown in Fig.~\ref{fig:pawn_PW_noSB_dep}. We confirm that this set-up explores a parameter space close to the second aforementioned family, where the temperature is lower and the mass loss rate is higher. In this case, the best fit is again better, and have a BIC=5014.8. The $\Delta{\rm BIC}=2.4$ from the hydrodynamical-Boltzmann model, meaning a small positive evidence against the LTE model (and only a smaller $\Delta{\rm BIC}=1.3$ from the hydrostatic Boltzmann model). In any case, the $\Delta{\rm BIC}$ is not sufficient to conclude in favor of any NLTE models (see the comparison in Fig~\ref{fig:best_fit_PW_SB}). Nevertheless it is interesting to note a few particularities in this scenario. First, the $\beta$ parameter is strongly correlated with the temperature and mass loss rate, giving a conservative idea of the uncertainties on these two parameters. Second, the mass loss rate is $10^{14.3^{+1.2}_{-1.4}}$~g\,s$^{-1}$ (between $8\times10^{12}$ and $3\times10^{15}$~g\,s$^{-1}$). Such a mass loss rate could have dramatic consequences (see discussion in Sec.~\ref{Sec_Discuss}). Third, the departure from LTE is $10^{-6.1^{+1.3}_{-1.1}}$ (in the $10^{-8.7}$--$10^{-2.2}$ range at 3$\sigma$), making the LTE excluded in this framework. Interestingly, our best value is far (but still compatible at $\sim$2.5~$\sigma$) from the theoretical work by \citet{GarciaMunoz2019} who predicted a departure from Boltzmann to be at the level of $\sim$$10^{-3}$. Our finding could hint to another possible departure from LTE than only the one from the Boltzmann equilibrium (see discussion in Sec.~\ref{Sec_Discuss}).

%%%%%%%%%%%%%%%%%%%%%%%%%%%%%%%%%%%%%%%%%%%%%%%%%%%%%%%%%%%%%%%
\section{Discussion}\label{Sec_Discuss}
%%%%%%%%%%%%%%%%%%%%%%%%%%%%%%%%%%%%%%%%%%%%%%%%%%%%%%%%%%%%%%%

\begin{figure}[t!]
\centering
\includegraphics[width=0.47\textwidth]{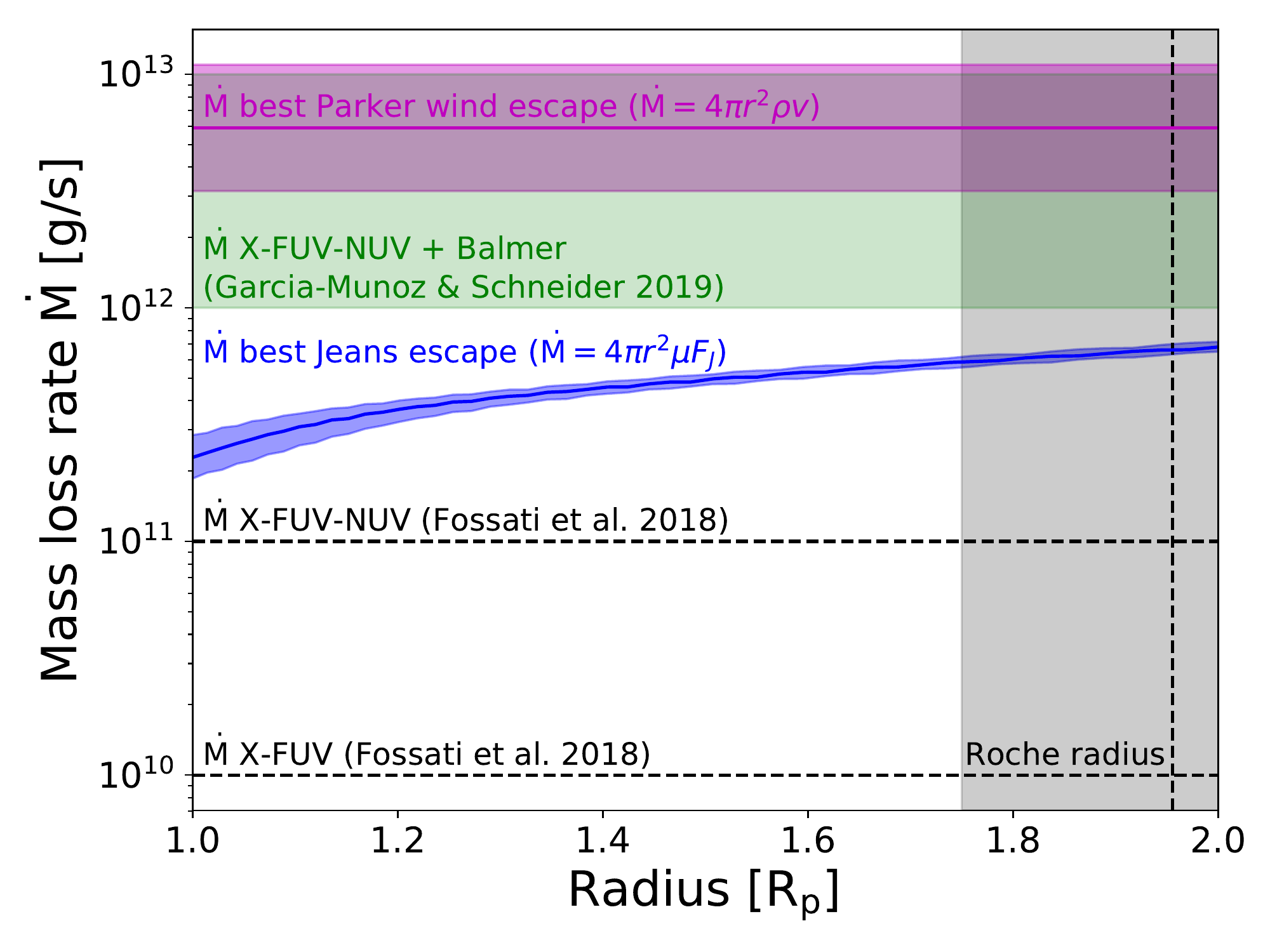}
\caption{Summary of the mass loss rates and their errors found in the LTE case compared to the expected mass loss rates from the theoretical studies of \citet{Fossati2018} and \citet{GarciaMunoz2019}. Our finding shows that the KELT-9~b thermosphere probably undergoes a ``Balmer-driven'' expansion, that leads to a high escape rate. In this case, the total mass loss through the planetary life time (500~$My$) represents $\sim$0.05~${\rm M_{Jup}}$, which is $\sim$2\% of the planet mass.}
\label{fig:MLR_summary}
\end{figure}

With our \texttt{PAWN} framework, we are able to retrieve a good fit to the data (Fig.~\ref{fig:best_fit_PW_SB}), and show that the thermosphere temperature (in LTE) is near 13\,000~K and the mass loss rate is around 10$^{12}$~g\,s$^{-1}$. If the hydrodynamic expansion of the thermosphere is a valid assumption, then the thermosphere temperature is constrained to be $T=13\,200^{+800}_{-720}$~K, and the mass loss rate to be $\dot{{\rm M}}_{\rm PW}=10^{12.8\pm0.3}$~g\,s$^{-1}$. This mass loss rate is higher than the energy-limited escape rate of 10$^{10}$--10$^{11}$~g\,s$^{-1}$ found by \citet[][see Fig.~\ref{fig:MLR_summary}]{Fossati2018}. It is important to note that this study only considers the X-FUV and NUV irradiation from the star, which may be small in the case of a A0-type star without chromosphere. However, \citet{GarciaMunoz2019} presented a new escape mechanism that applies to ultra-hot Jupiters orbiting early-type stars. This mechanism shows that it is important to also consider the NUV and optical Balmer continuum and Balmer series irradiation from the star. In this scenario, the thermosphere is first heated by Lyman photons (Lyman continuum and especially Ly$\alpha$), which increases the number density of excited hydrogen in the high atmosphere (Lyman irradiation and de-excitation put the population levels out of Boltzmann equilibrium), thus increasing the Balmer opacity. At a certain point, the excited hydrogen atoms are able to absorb significantly much more flux from the stellar Balmer photons, which constitutes an enormous source of energy in early-type stars. This increases even more the thermosphere temperature, which reaches an equilibrium temperature of $\sim$12\,000--15\,000~K (3\,000~K more than with only the X-FUV-NUV flux). In this case the mass loss rate increases to 10$^{12}$--10$^{13}$~g\,s$^{-1}$, the atmosphere undergoes a ``Balmer-driven'' escape, named so because the Balmer energy deposition dominates over all other. This escape rate is well in line with our findings (Fig.~\ref{fig:MLR_summary}). Knowing that the age of KELT-9 is $\sim$300--500~My \citep{Gaudi2017,Borsa2019}, a mass loss of this rate over the planetary life time represents a total mass loss of $\sim$0.03--0.05~${\rm M_{Jup}}$ ($\sim$1.5--2\% of the planet mass).

In the thermosphere, the excited hydrogen atoms are expected to be in departure from LTE. The main reason being the lower frequency of collisions at high altitudes, making the hydrogen energy levels depart from the Boltzmann equilibrium, typically with a $\beta$ factor of 10$^{-3}$ \citep{GarciaMunoz2019}. Despite the data being too low in S/N to show clear evidences for a departure from the LTE, our NLTE model hints at a lower temperature of $T=9\,600\pm1200$~K, and a higher mass loss rate of $\log_{10}({\rm \dot{M}}_{\rm PW}\,[{\rm g\,s^{-1}}])=14.3^{+1.2}_{-1.4}$. If true, the planet may have already lost 1--2~${\rm M_{Jup}}$, and would undergo a catastrophic evaporation that would make the planet barely survive the next $\sim$500--700~My, when its host star will enter the red-giant phase. Such a mass loss rate is not yet physically motivated \citep[though see][]{GarciaMunoz2019}. One possibility would be to explore the real effect of the stellar gravity and magnetic field. Indeed, such effects could push down the sonic point to much lower altitude, forcing the density and velocity to increase at high altitude, hence increasing the mass loss rate \citep{MurrayClay2009}. Another possibility would be to investigate the effect of the high rotation of KELT-9, which is also relatively young. Indeed, both effects potentially increase the mass loss rate \citep{Allan2019}.

In our NLTE model, we found that the departure coefficient $\beta$ is on the order of 10$^{-6}$, while theory predicts 10$^{-3}$ (only for Boltzmann departure). Thus, our much smaller $\beta$ coefficient could hint to some other possible LTE departure. Indeed, because $\beta$ is eventually normalized to the total amount of neutral hydrogen, our finding may signify that there is less neutral hydrogen than in the chemical (Saha) equilibrium by a factor of 10$^3$. This implies that the LTE departure may also be caused by photo-ionization \citep[Saha equilibrium departure, e.g.,][]{Fortney2003,Sing2008b,VidalMadjar2011b,Lavvas2014}. In this case, the ratio of neutral to ionized hydrogen would be 10$^{-3}$ lower. The other atoms would probably also undergo photo-ionization, making the ionized species even more dominant in the high atmosphere, increasing the importance of ionospheric phenomenons \citep{Helling2019}.

\begin{figure}[t!]
\centering
\includegraphics[width=0.47\textwidth]{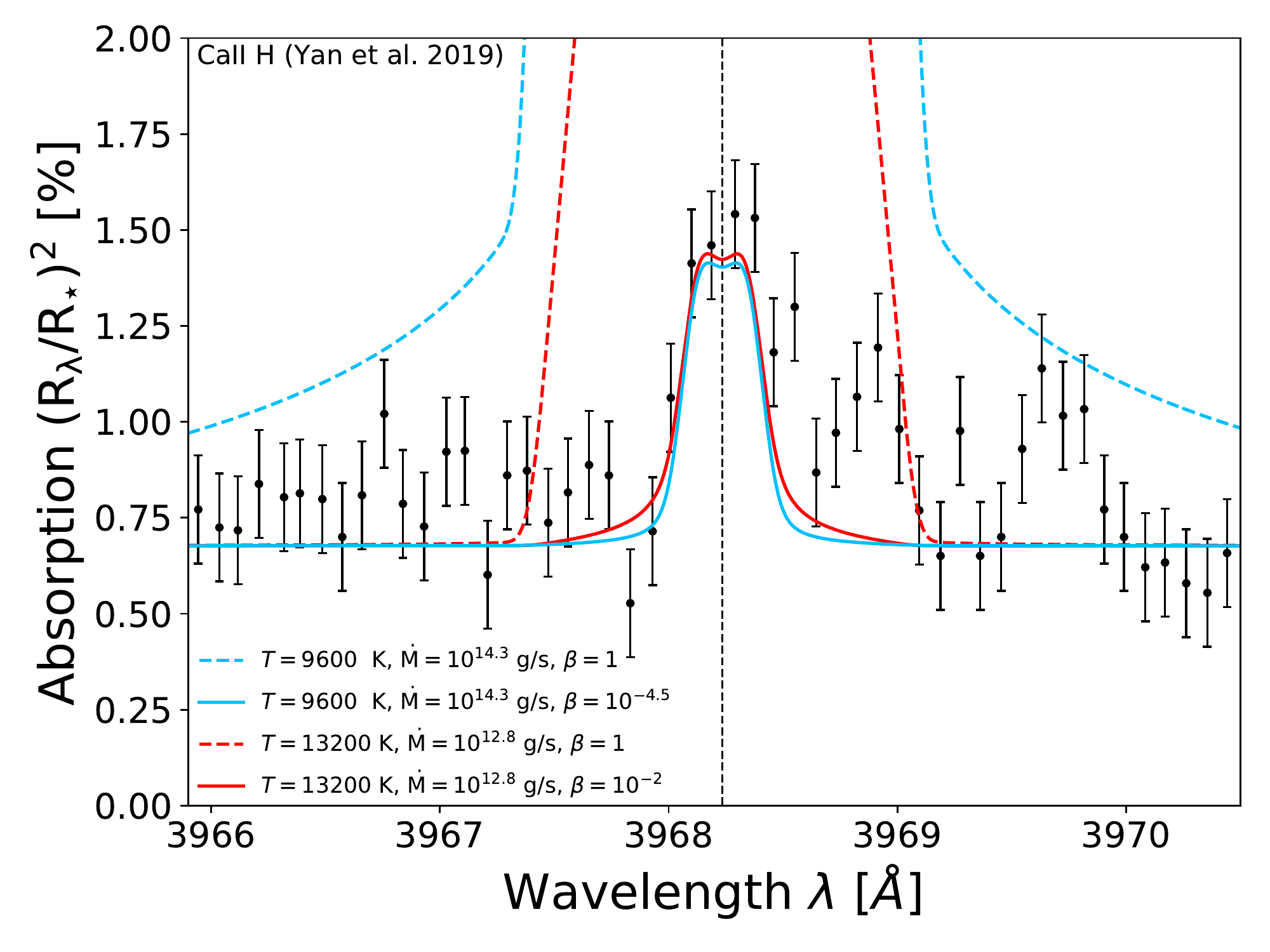}
\caption{Example of \texttt{PAWN} models of the \ion{Ca}{ii} H line detection by \citet{Yan2019}. Our model is compatible with their detection only when assuming some LTE departure. For the \ion{Ca}{ii}, the departure is likely explained by further thermal and/or photo-ionization not yet taken in our code into account. Additional data gathering, and modeling will be necessary to confirm our interpretation.}
\label{fig:calcium}
\end{figure}

Moreover, our NLTE scenario may be able to explain the strong absorption by ionized atomic species, especially the \ion{Ca}{ii} and \ion{Fe}{ii}, by \citet{Hoeijmakers2018,Hoeijmakers2019,Cauley2018,Yan2019,Turner2020}. It is interesting that \citet{Yan2019} found a different line ratio between the calcium doublet and triplet for the two ultra-hot Jupiter WASP-33~b and KELT-9~b. Hence, this type of data is also sensing the local thermodynamical conditions, in complement to the Balmer lines. We can predict absorptions of the calcium doublet \ion{Ca}{ii} H\&K, and infrared triplet (IRT) lines, by taking the two main atmospheric structures of our best fit models of the Balmer lines. To model the calcium lines, we assume the calcium abundance to be solar \citep[$\chi_ {Ca}$=6.34,][]{Asplund2009}, and ${\rm r_{up}}$=10 (for good convergence). In our code, the calcium population is dominated by ionized \ion{Ca}{ii} at the temperatures and pressures of our models \citep[see also][although for very high temperatures, the \ion{Ca}{iii} and further ionization levels are becoming important]{Kitzmann2018,Lothringer2018,Yan2019}. In this case, our model is compatible with the detections of \citet{Yan2019,Turner2020} only if we assume a departure from LTE on the order of $\sim$10$^{-2}$, or $\sim$10$^{-4.5}$ for our two main structures, respectively (Fig~\ref{fig:calcium}). Furthermore, the \ion{Ca}{ii} opacities are much larger than the Balmer opacities (by $\sim$100 times), but the measured absorptions have about the same contrasts, which hints at the presence of some \ion{Ca}{ii} NLTE. Because the calcium H\&K lines are arising from the ground state of the singly ionized calcium, Boltzmann equilibrium departure is not likely to explain this, because the lack of collision is mostly populating the ground state level (the smaller calcium IRT absorption may be explained by non-Boltzmann distribution since this triplet arises from the first excited level). The $\beta$ departure in the \ion{Ca}{ii} case may also be due to a non-solar metallicity, but it is unlikely that the planet atmosphere metallicity is far below the solar one (KELT-9 has a solar metallicity, and planets are expected to generally have a higher metallicity than their host star). Thus, the departure from LTE may be better explained by a second ionization of calcium (through thermal and photo-ionization), which would lower the singularly ionized calcium population. The departure level for \ion{Ca}{ii} is in line with the extra departure found for hydrogen (we note that the ionization energy of the \ion{H}{i} and \ion{Ca}{ii} have similar value of 13.6~eV and 11.9~eV, respectively). More investigations have to be done to explore if the hypothesis of chemical equilibrium in the thermosphere of (ultra) hot Jupiter is still justified.

One caveat of our framework is that our set of assumptions may be too strong when interpreting, together or not, the hydrogen and calcium absorptions. For example, the assumption of an isothermal atmosphere is probably too strong because temperature gradients and temperature inversion are predicted \citep{Kitzmann2018,Lothringer2018,Mansfield2019}. Actually, such a temperature inversion has been detected for the first time by \citet{Pino2020} in the KELT-9~b atmosphere by observing neutral iron lines in emission on the day side. Their detection shows that the temperature inversion happens from the 1~bar region up to the 10$^{-3}$--10$^{-5}$ bar region. However, \citet{Hoeijmakers2019} infers that the maximum probable pressure in transmission is in the mbar regime, due to the H$^-$ opacity in the optical. The consequence for our model is that defining the lower boundary for the radius integration at 1~${\rm R_P}$ (the measured white light radius) is justified because this radius is set by the, effectively gray, H$^-$ opacity. At no wavelength should the high-resolution observations probe smaller radii. However, because we assume that the atmosphere is isothermal, at high temperatures this can lead to pressures at 1~${\rm R_P}$ that are much smaller than mbar. Forcing a fit to continue to mbar pressures increases the line core to continuum amplitude unrealistically. Such a continuation to larger pressures at fixed temperature is also not correct because the atmosphere must transition from the high thermosphere temperature to the planetary equilibrium temperature. Accounting for the decrease in atmospheric scale-height, the line core to continuum amplitude would then stay compatible with our current model. Another effect, not considered in our model, is a potential difference in the vertical distribution of the individual atmospheric gases that are determined by their respective masses, whose effect is to give birth to a heterosphere on top of the homosphere \citep{Shizgal1996}. For example, the presence of a heterosphere may also explain the weak calcium absorption, since these atoms would stay in the lower thermosphere. However, this effect of atomic diffusion may be counterbalanced by the hydrodynamic outflow that is able to lift heavy atomic species, that has been observed in exospheres \citep{VidalMadjar2004}. A last caveat is the modeling of the exosphere itself. Because the Roche radius of KELT-9~b is low ($1.95\pm0.2~{\rm R_P}$), it is expected that hydrogen and metals escape easily. The kinetic broadening in the exosphere would change the shape of the lines, making them asymmetric \citep{Ehrenreich2015,Bourrier2015,Bourrier2016a,Allart2019}. This effect has not yet been seen for the KELT-9~b Balmer lines. Thus, additional data gathering, and modeling will be necessary to get a coherent image of the KELT-9~b atmosphere consistent with all the observations.

%%%%%%%%%%%%%%%%%%%%%%%%%%%%%%%%%%%%%%%%%%%%%%%%%%%%%%%%%%%%%%%
\section{Conclusion}\label{Sec_Conclu}
%%%%%%%%%%%%%%%%%%%%%%%%%%%%%%%%%%%%%%%%%%%%%%%%%%%%%%%%%%%%%%%

We analyzed two optical high-resolution transit observations of KELT-9~b gathered with HARPS-N, with our \texttt{CHESS} framework. We found a spin-orbit misalignment of $\lambda=\ang{-85.01}\pm\ang{0.23}$, and positive evidence in favor of a differential rotation model. We also confirmed the presence of stellar pulsations with a period of ${P_{\rm puls}}=7.54\pm0.12$~h. We use the reloaded RM analysis to correct the Balmer lines from the local stellar photospheric spectrum effect in transmission spectroscopy. We detected the \Ha, \Hb, \Hg, and \Hd\ absorptions coming from the planet's upper atmosphere, with high confidence (between 24.3 and 4.6~$\sigma$) during both nights. The \Ha, \Hb, \Hg\ lines probe altitudes up to $\sim$1.2 to $\sim$1.5 planetary radii; they do not show any significant velocity shifts, and share a large FWHM of $44.9\pm1.4$~\kms\ that is dominated by optically thick absorptions. We presented a new model, \texttt{CHESS.PAWN}, to interpret the absorption in the Balmer series in exoplanet atmospheres that is linked with a retrieval algorithm. We constrained the thermosphere temperature to be $T=13\,200^{+800}_{-720}$~K, and the mass loss rate to be $\dot{{\rm M}}_{\rm PW}=10^{12.8\pm0.3}$~g\,s$^{-1}$, under the assumptions of a hydrodynamic expansion and of LTE. This escape rate is fully explained by the ``Balmer-driven'' escape mechanism proposed by \citet{GarciaMunoz2019}. NLTE effects are not yet significantly detected with this data set.

Our work shows that with the infrared metastable Helium triplet \citep[$\lambda 10830\AA$,][]{Seager2000,Oklopcic2018,Spake2018,Allart2018,Nortmann2018}, the Balmer series is an alternative new tool for probing exoplanet evaporation, alongside with classical UV observations \citep{VidalMadjar2003}. Furthermore, these two probes seems to be complementary since the helium lines are mainly sensitive to G and K-type exoplanetary systems \citep{Oklopcic2019}, while the Balmer lines are more sensitive to early-type (A-type to G-type) systems \citep{Jensen2012,Jensen2018,Yan2018,Casasayas-Barris2018,Casasayas-Barris2019,Cabot2020,Chen2020}. Finally, such observations will benefit from the ESPRESSO spectrograph \citep{Pepe2013,Pepe2014}. On the one hand, ESPRESSO will increase the S/N of the absorption lines \citep{Ehrenreich2020,Chen2020}. On the other hand, in combination with near-infrared spectrographs (CARMENES, SPIRou, NIRPS), it will be possible to detect more absorption lines arising from different electronic levels of a given atomic species. For example, the combination of the Balmer and Paschen (or Brackett) series for the hydrogen, of the infrared triplet, and of the D3 line for the helium, or of the H\&K doublet and of the infrared triplet for the calcium, will allow us to infer the local thermodynamic state. This will eventually allow us to put constraints on several chemical and physical aeronomic processes.

%%%%%%%%%%%%%%%%%%%%%%%%%%%%%%%%%%%%%%%%%%%%%%%%%%%%%%%%%%%%%%%

\begin{acknowledgements}
A.W. acknowledges the financial support of the SNSF by grant number P2GEP2\_178191 and P400P2\_186765. This work has been carried out within the frame of the National Centre for Competence in Research ``PlanetS'' supported by the Swiss National Science Foundation (SNSF). This project has received funding from the European Research Council (ERC) under the European Union's Horizon 2020 research and innovation program (grant agreement number 724427; project Four Aces). I.S. and P.M acknowledges funding from the European Research Council (ERC) under the European Union's Horizon 2020 research and innovation program (grant agreement No. 694513 and 832428; ExoplanetBio). L.P. acknowledges that the research leading to these results has received funding from the European Research Council (ERC) under the European Union's Horizon 2020 research and innovation program (grant agreement No. 679633; Exo-Atmos). We thank the anonymous referee for the reading and the pertinent comments and questions. A.W. thank J. Alonso-Floriano for discussions about transmission spectroscopy, C. Aerts for discussions about stellar pulsations, and A. Vidotto, A. Garc\`ia-Mu\~noz, M. Bergemann, and Y. Miguel for discussions about NLTE in hot Jupiter atmospheres.
\end{acknowledgements}

\bibliographystyle{aa}

\clearpage
\begin{appendix}

\section{Rossiter-McLaughlin MCMC results}\label{app:MCMC_RM}

Here, we show the correlation diagrams for the reloaded Rossiter-McLaughlin study obtained with the \texttt{emcee} algorithm.

\begin{figure}[htbp]
\centering
\includegraphics[width=0.47\textwidth]{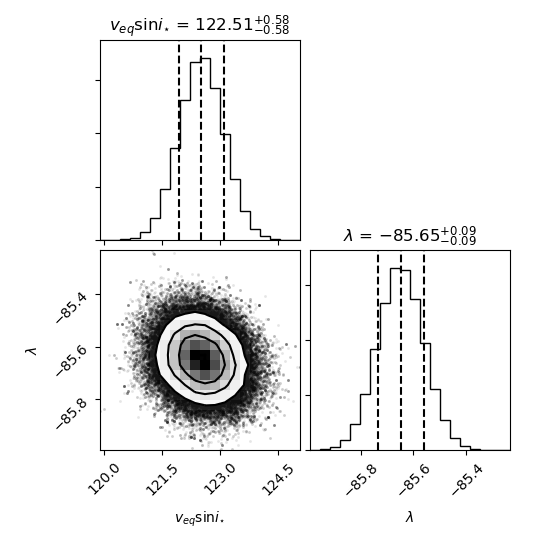}
\caption{Correlation diagrams in the case of a stellar solid body rotation.}
\label{fig:RM_SB}
\end{figure}

\begin{figure}[htbp]
\centering
\includegraphics[width=0.47\textwidth]{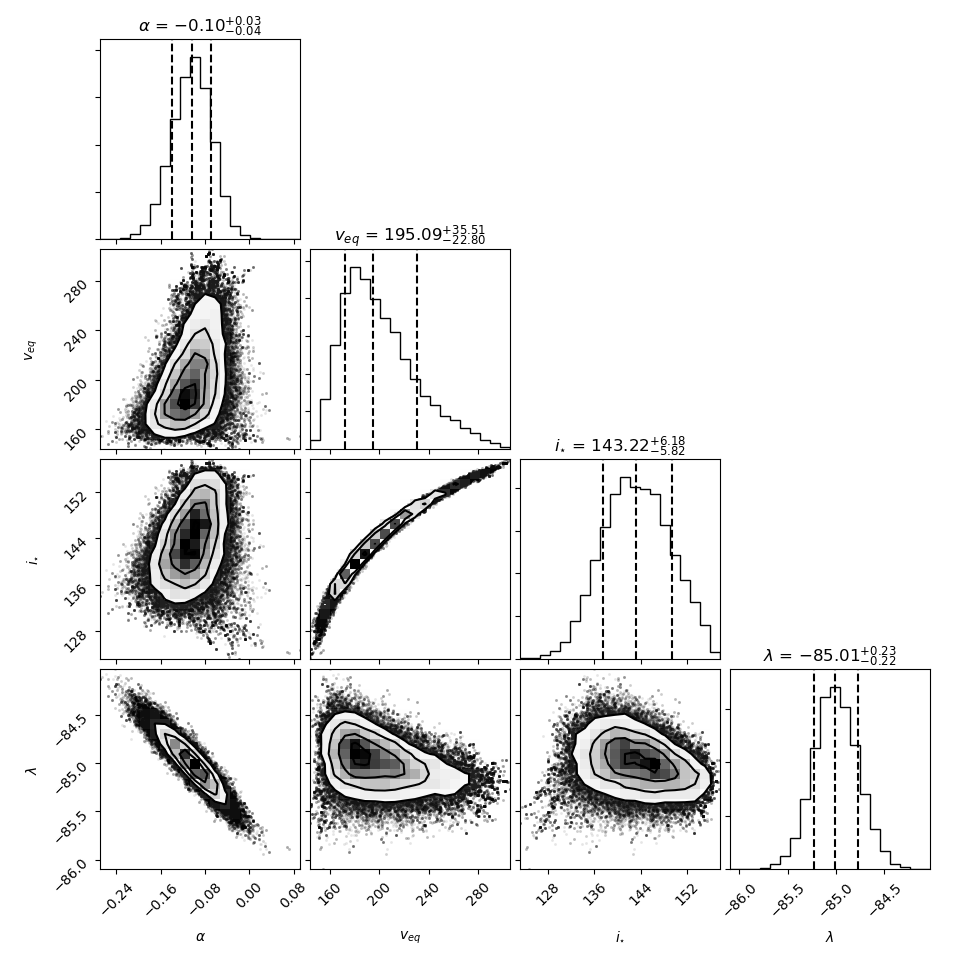}
\caption{Correlation diagrams for the case of a stellar differential rotation, with an ``anti-solar'' law.}
\label{fig:RM_DR_neg}
\end{figure}

\begin{figure}[htbp]
\centering
\includegraphics[width=0.47\textwidth]{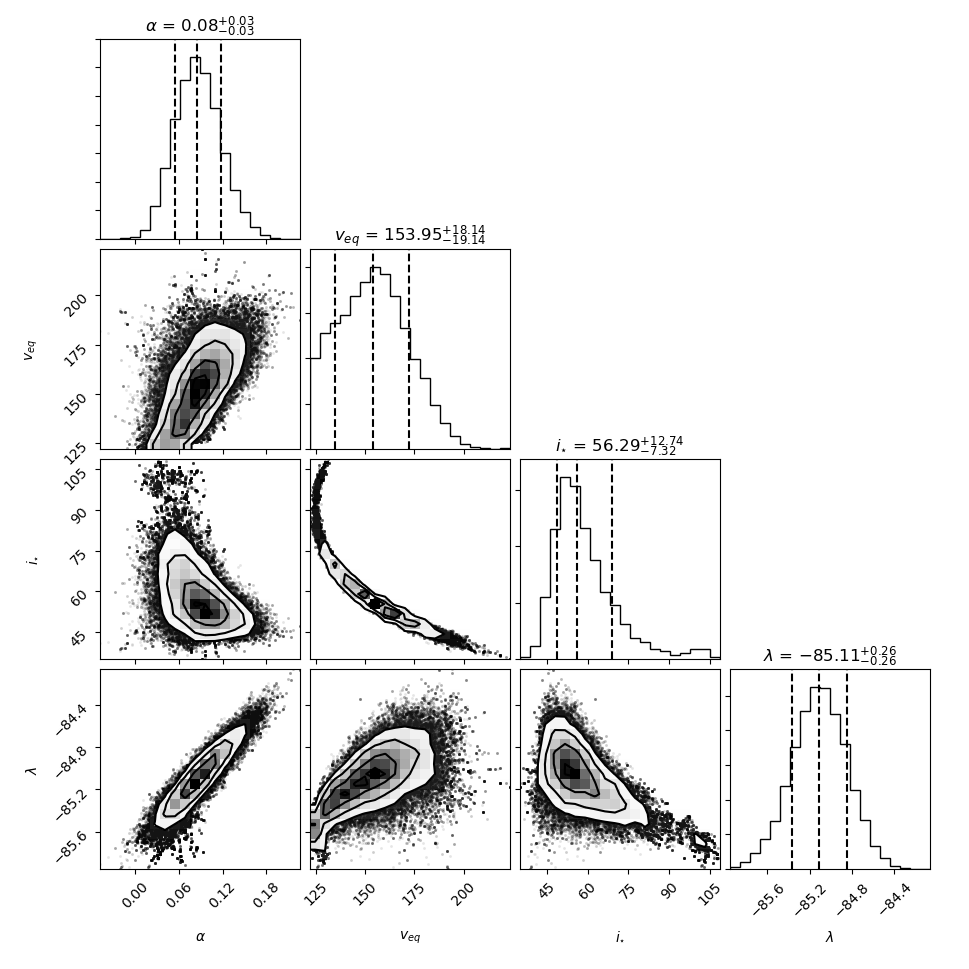}
\caption{Correlation diagrams for the case of a stellar differential rotation, with a solar-like law.}
\label{fig:RM_DR_pos}
\end{figure}

\section{Rossiter-McLaughlin correction for transmission spectroscopy}\label{app:RM_TS}

Here, we show the effect of the Rossiter-McLaughlin correction on the \Ha\ line presented in Sec.~\ref{Sec_TS}.

\begin{figure}[htbp]
\centering
\includegraphics[width=0.47\textwidth]{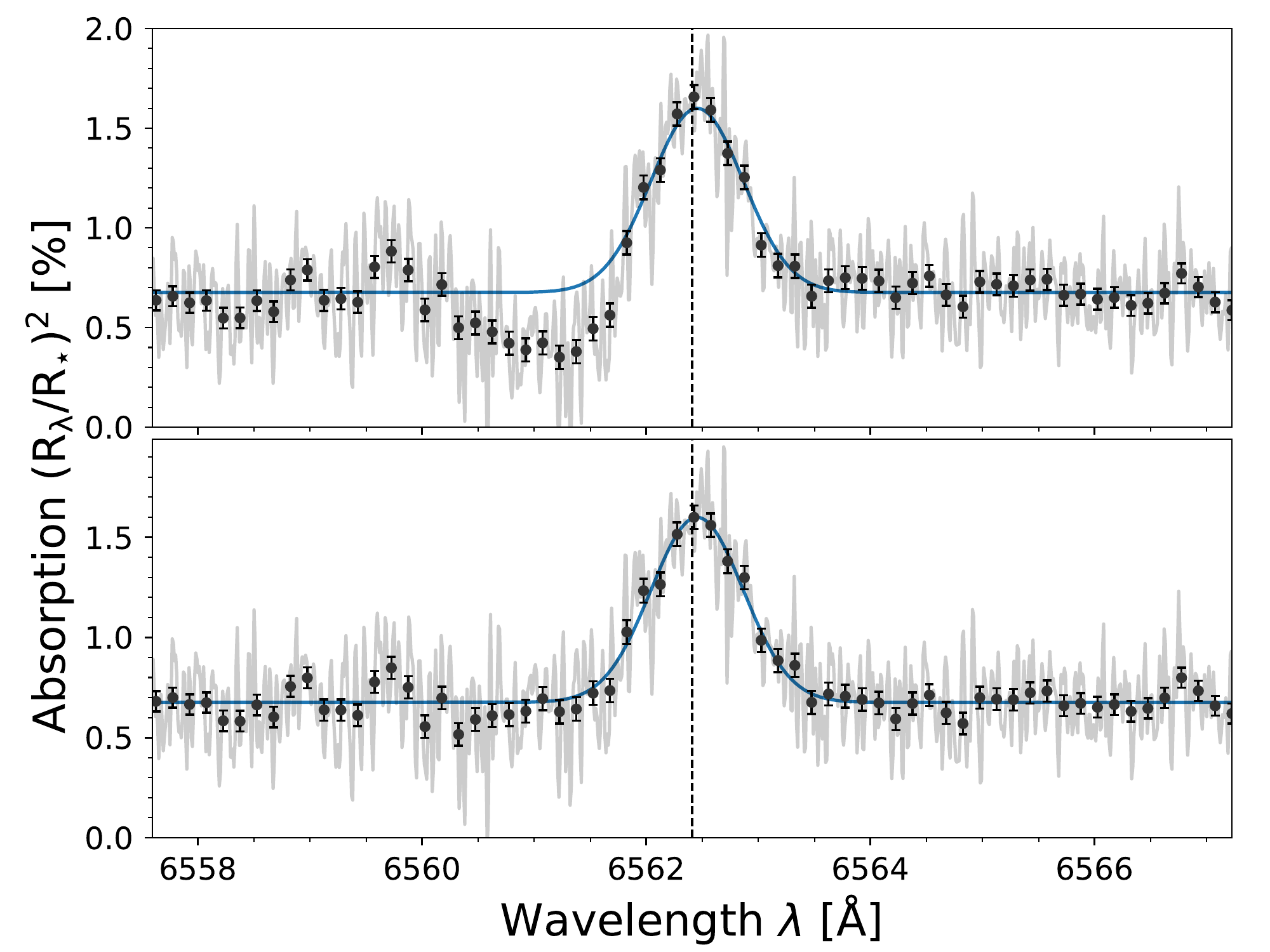}
\caption{\Ha\ transmission spectrum in the planet rest frame (gray line and black circles), uncorrected (top) and corrected (bottom) from the Rossiter-McLaughlin effect. The vertical black dashed line shows the line center taking the systemic velocity into account. We show the Gaussian fit (blue) to the corrected line in both sub-panels for the purposes of comparison.}
\label{fig:RM_TS_Ha}
\end{figure}

\section{Transmission spectroscopy MCMC results}\label{app:MCMC_TS}

Here, we show complementary correlation diagrams for the Balmer lines NLTE study presented in Sec.~\ref{sec:NLTE}.

\begin{figure*}[htbp]
\centering
\includegraphics[width=0.90\textwidth]{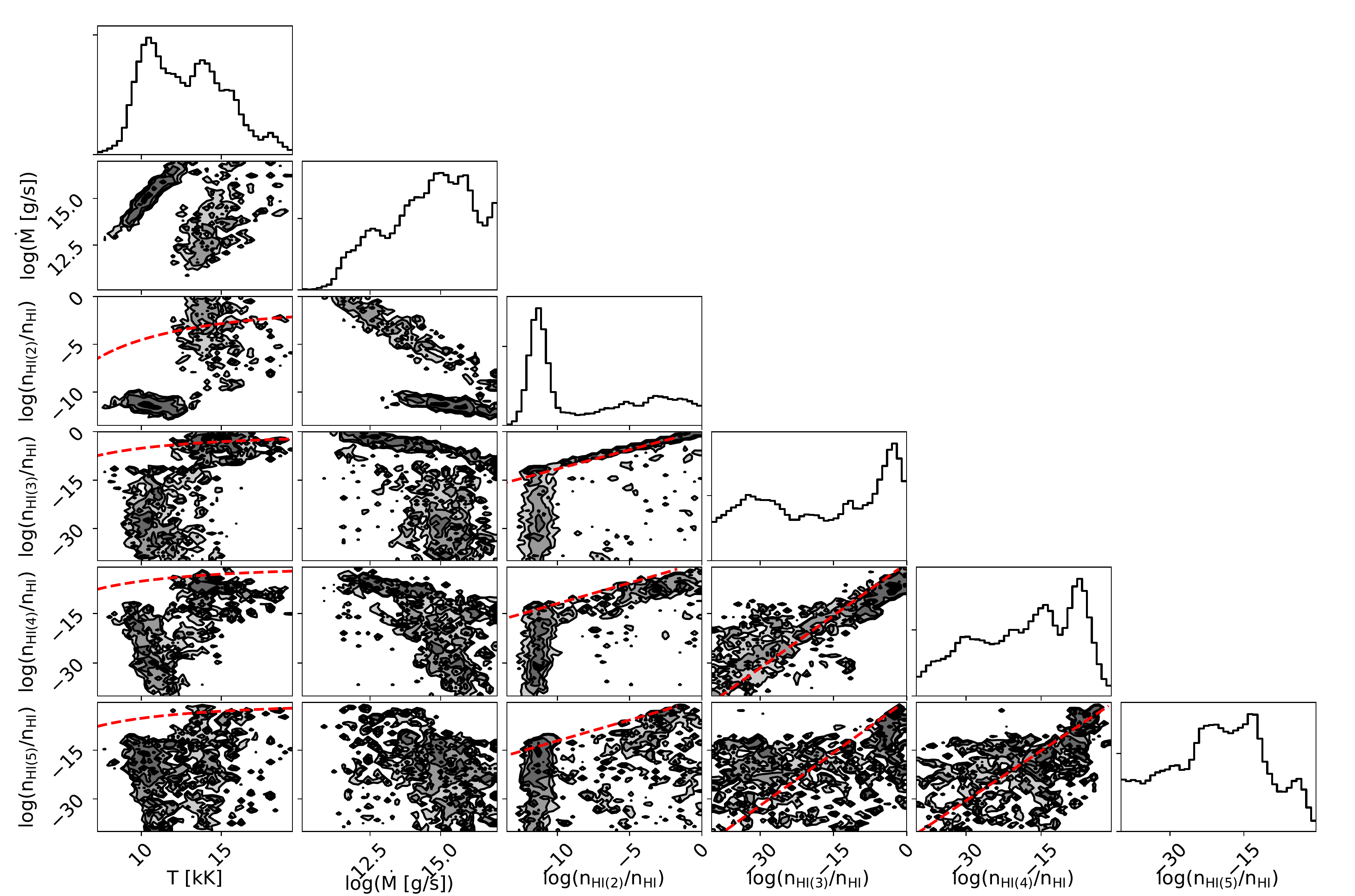}
\caption{Correlation diagrams in the case of a complete NLTE atmosphere (each neutral hydrogen electronic level is a separated free parameter). Two families of solution are visible, one being compatible and not so far from the Boltzmann equilibrium (shown in red dashed lines), the other being significantly away from the Boltzmann equilibrium. These two families are studied in the main text.}
\label{fig:pawn_PW_noSB_free_all}
\end{figure*}

\begin{figure*}[t!]
\centering
\includegraphics[width=0.90\textwidth]{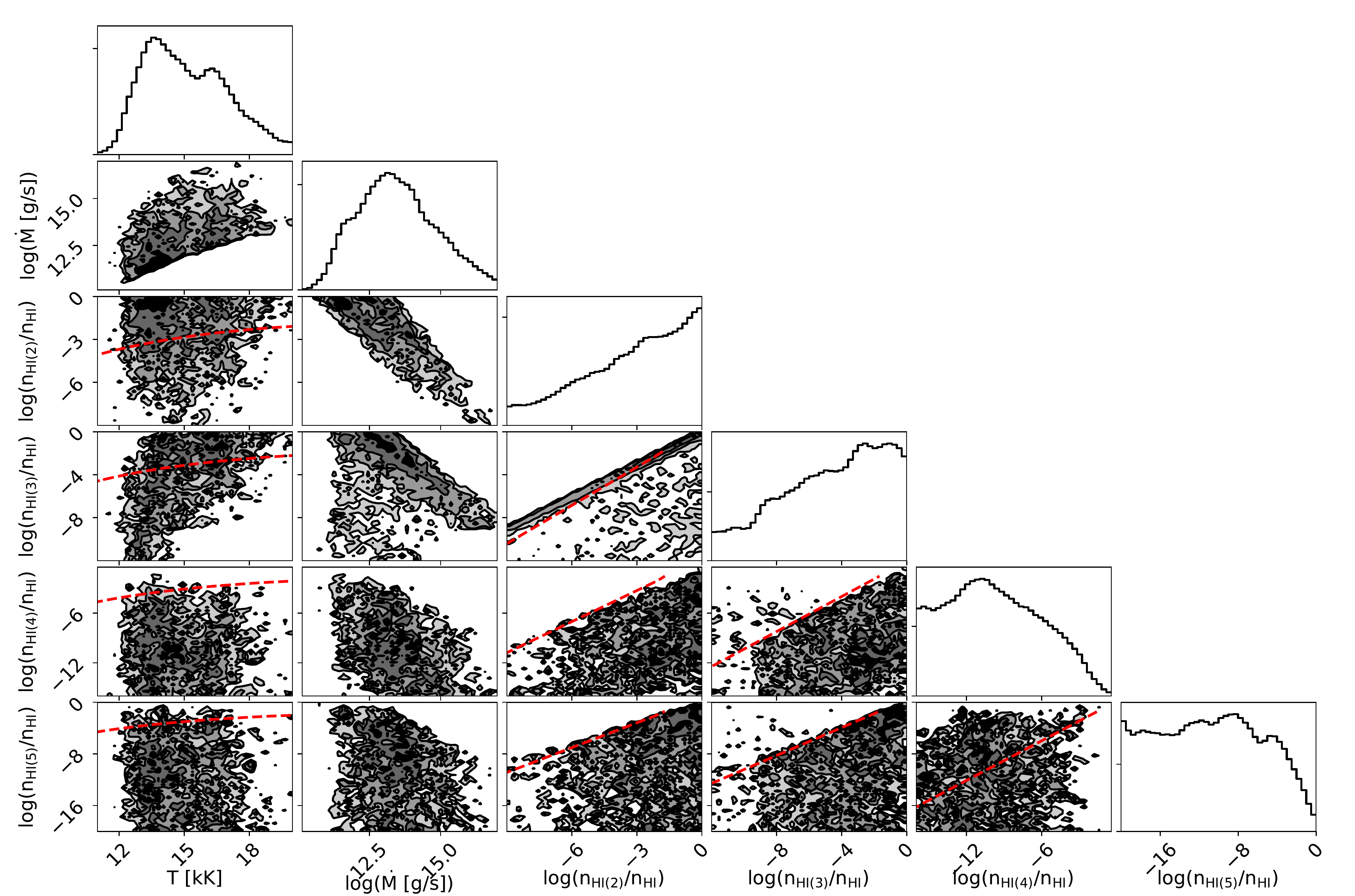}
\caption{Correlation diagrams in the case of a complete NLTE atmosphere (each neutral hydrogen electronic level is a separated free parameter). Exploration of the high temperature, near Boltzmann equilibrium (red dashed lines) family solution (see main text).}
\label{fig:pawn_PW_noSB_free}
\end{figure*}

\end{appendix}

\end{document}